\documentclass{aastex63}
\usepackage{physics}
\usepackage{savesym}
\savesymbol{tablenum}
\usepackage{siunitx}
\restoresymbol{SIX}{tablenum}
\usepackage{dsfont}
\usepackage{color}

\let\vec\boldsymbol

\submitjournal{The Astrophysical Journal}

\shorttitle{Line-tied BCs generate unphysical boundary layers}
\shortauthors{A.~P.~K.~Prokopyszyn et al.}

\begin{document}

\title{Line-tied boundary conditions can cause resonant absorption models to generate unphysically large boundary layers}

\correspondingauthor{A.~P.~K.~Prokopyszyn}
\email{apkp@st-andrews.ac.uk}

\author[0000-0002-8184-5990]{A.~P.~K.~Prokopyszyn}
\affiliation{School of Mathematics and Statistics, University of St Andrews, St Andrews, Fife, KY16 9SS, U.K.}

\author[0000-0002-9877-1457]{A.~N.~Wright}
\affiliation{School of Mathematics and Statistics, University of St Andrews, St Andrews, Fife, KY16 9SS, U.K.}

\author[0000-0003-2620-2068]{A.~W.~Hood}
\affiliation{School of Mathematics and Statistics, University of St Andrews, St Andrews, Fife, KY16 9SS, U.K.}

\begin{abstract}

This paper uses linear magnetohydrodynamics to model resonant absorption in coronal plasma with a Cartesian coordinate system. We impose line-tied boundary conditions and tilt the background magnetic field to be oblique to the transition region. \citet{Halberstadt1993,Halberstadt1995,Goedbloed1994,Arregui2003} show that line-tied boundary conditions cause their resonant absorption models to produce steep boundary layers/evanescent fast waves. We aim to study the importance of the boundary layers and assess their significance in a solar context. We calculate the solutions in a model where we impose line-tied boundary conditions and compare this with a model where we include the chromosphere instead. Results are calculated analytically and then verified numerically. We show that line-tied boundary conditions can cause the model to overestimate the boundary layers' amplitude significantly. If the fast waves can propagate in the chromosphere, then the line-tied model accurately predicts the boundary layers' amplitude. However, if the fast waves are evanescent, then the boundary layers' size is reduced significantly, and the line-tied model overestimates their amplitude. This leads to the counter-intuitive result that the length scales tangential to the transition region can play an essential role in determining line-tied boundary conditions' validity. The results suggest that line-tied boundary conditions can cause the model to generate unphysically large boundary layers. However, researchers may wish to continue to use them in their models for their simplicity and ability to significantly reduce computation time if they understand and are aware of their flaws.

\end{abstract}

\section{Introduction}

Magnetohydrodynamic (MHD) waves are commonplace in the solar atmosphere. They have been observed over the last two decades as a consequence of new, improved imaging and spectroscopic instruments (see, for example, \citet{Tomczyk2007}, \citet{McIntosh2011} and \citet{DeMoortel2012}). \citet{Goossens2011} gives a review of the linear behaviour of MHD waves. The dissipation of Alfv\'en waves has been the basis of many coronal heating models (see review by \citet{Arregui2015} and references therein). 

This paper focuses on studying resonant absorption (see, for example, \citealt{Ionson1982}) which describes the process where magnetoacoustic waves with a given frequency mode convert to Alfv\'en / Alfv\'enic waves at a location where the background magnetic field lines have the same natural Alfv\'en frequency. If the magnetoacoustic waves' frequency equals the field lines' natural frequency, then resonance is produced. Here mode conversion can convert a magnetoacoustic wave into an Alfv\'en wave. This is a one-way process that results in a localised, high concentration of energy, in a process which is analogous to Barton's pendulum experiment. Barton's investigation shows a heavy pendulum driving a series of lighter pendula/oscillators. One of these lighter oscillators, called the resonant pendulum, has the same natural frequency as the driving frequency. Since all the oscillators connect to the same string, they are all driven at the same frequency. After a while, the system reaches a steady-state where the whole system oscillates at the driver frequency. At steady- state, the resonant pendulum oscillates at a significantly larger amplitude than the other oscillators. The string which connects the oscillators is analogous to magnetoacoustic waves, and the oscillators are comparable to standing Alfv\'en waves. For both cases, energy concentrates in a narrow region due to resonance.

Resonant absorption plays a crucial role in coronal seismology. Coronal seismology is becoming an increasingly relevant field due to developments in observational instruments, computational power and MHD wave theory \citep{DeMoortel2012}. It is a technique where mathematical models are used to infer hard to measure quantities, e.g. the coronal magnetic field strength or density gradients from easier to measure quantities, e.g. wave frequencies and damping time. Advances in coronal seismology could provide us with better knowledge of physical values in the solar atmosphere. This may help us solve problems such as the coronal heating problem, and it could present us with new puzzles to solve. Resonant absorption can play a vital role in damping observed coronal loop oscillations \citep{Nakariakov1999,Terradas2006}. \citet{Ruderman2002} show that a kink wave in a cylindrical flux tube can mode convert to a torsional Alfv\'en wave. This results in the kink wave amplitude following a decay profile that closely resembles observed decay profiles. Resonant absorption can also play a key role in triggering the Kelvin-Helmholtz instability in coronal loops which causes the plasma flow to become turbulent (see, e.g. \citealt{Antolin2016,Howson2017}).

In our models, we drive fast magnetoacoustic waves in from the side for mathematical convenience. These waves could be generated by, for example, a nearby flare. However, it is not our aim to investigate how the waves enter the corona, but instead, we study their dynamics
once they are in the corona. The origin of coronal waves remains an open question. It is difficult for Alfv\'en waves generated at the photosphere to enter the corona \citep{Cranmer2005}, due to the rapid exponential decay of the density with height in the chromosphere and the steep decrease at the transition region. \citet{Hollweg1984} suggests resonances in coronal loops and spicules provide enough energy flux to the corona to match the observed wave velocity amplitudes. \citet{Cally2011,Hansen2012} suggest that mode conversion from fast waves to Alfv\'en waves at the transition region enables sufficient energy flux to enter the corona. It is also possible that the corona itself generates Alfv\'en waves via magnetic reconnection \citep{Cranmer2018} and the energy released during reconnection partially feeds into the energy contained in the Alfv\'en waves.

This paper closely follows the seminal work of \citet{Halberstadt1993,Goedbloed1994,Halberstadt1995}. They modelled linear MHD waves and resonant absorption in a closed domain where the background magnetic field is tilted to be oblique to the solar surface. They showed that imposing line-tied boundary conditions causes the Alfv\'en waves and fast waves to couple. Suppose the transverse length scales are shorter than the Alfv\'en wavelength then evanescent fast waves/boundary layers from at the line-tied boundaries. Note that resonant absorption typically generates transverse length scales, which are many times shorter than the Alfv\'en wavelength. \citet{Goedbloed1994} clearly shows the boundary layers in Figure 6c of their paper for the fast wave components of the Lagrangian displacements ($\xi_F$ and $\eta_F$). We recommend reading \citet{Goedbloed1994} for an in-depth explanation of how the boundary layers form, but we also give a brief explanation in the penultimate paragraph of Section \ref{sec:uniform_alfven_speed}. \citet{Arregui2003} confirms the existence of these boundary layers via a numerical approach. One of the key differences between the work presented in this paper to that already in the literature is that we test the validity of imposing line-tied boundary conditions. To test the validity, we check if a model which includes the chromosphere instead of imposing line-tied boundary conditions can produce similar boundary layers. We model the corona and chromosphere using a piecewise constant background density profile with a discontinuous jump from the corona to the chromosphere.


This paper's outline is as follows; in Section \ref{sec:model_and_equations} we present the equations we will use to model the plasma and then discuss the applicability of some of the assumptions we make in the corona. In Section \ref{sec:uniform_alfven_speed}, we calculate normal mode solutions in a domain where the Alfv\'en speed is uniform. We impose an incident Alfv\'en wave and calculate the unique solution which ensures the velocity is zero at $z=0$. Since the Alfv\'en speed is uniform, resonant absorption cannot occur. However, we choose the length scales in $x$ to be extremely short to simulate the conditions near a singularity in a resonant absorption experiment. Section \ref{sec:piecewise_constant_alfven_speed} builds on Section \ref{sec:uniform_alfven_speed} by using a background Alf\'en speed which is piecewise constant in $z$. The regions $z>0$ and $z<0$ corresponds to the corona and chromosphere, respectively. We impose an incident Alfv\'en wave and calculate the unique solution which ensures continuity of the velocity and its derivative at $z=0$. By comparing results from Sections \ref{sec:uniform_alfven_speed} and \ref{sec:piecewise_constant_alfven_speed}, we can test how robust the boundary layers produced by line-tied boundary conditions are when we consider the finite nature of the jump in Alfv\'en speeds and the small length scales of a resonance layer. In Section \ref{sec:resonant_absorption_model} we use a background Alfv\'en speed which is a function of $x$ and piecewise constant in $z$. Our goal here is to test if steep boundary layers form in a resonant absorption experiment. Finally, in Section \ref{sec:summary}, a summary of our results and conclusions are given.

\section{Model and equations}
\label{sec:model_and_equations}

In this section, we describe the most general model we will use. However, throughout this paper, we will start with a simple model. In subsequent sections, we will add complexity to the model as we head
towards the most relevant description of resonance in coronal magnetic fields. For example, in Sections \ref{sec:uniform_alfven_speed} and \ref{sec:piecewise_constant_alfven_speed}, the background Alfv\'en speed, $v_A$, will not depend on $x$ but, in Section \ref{sec:resonant_absorption_model}, we include a variation in $x$ as well as a model chromosphere. We consider linear perturbations on a static background equilibrium. Thus, we assume that
\begin{eqnarray}
    \label{eq:linearisation_assumption1}
    u(\vec{x},t) & \ll & v_A, \\
    \label{eq:linearisation_assumption2}
    \vec{B}(\vec{x},t) & = & \vec{B}_0(\vec{x}) + \vec{b}(\vec{x},t),\ \text{where}\ b \ll B_0, \\
    \label{eq:linearisation_assumption3}
    \rho(\vec{x},t) & = & \rho_0(\vec{x}) + \rho_1(\vec{x},t),\ \text{where}\ \rho_1 \ll \rho_0,
\end{eqnarray}
and 
\begin{equation}
    u = \abs{\vec{u}},\ B_0=\abs{\vec{B}_0},\ b=\abs{\vec{b}},
\end{equation}
where $\vec{u}$ denotes the plasma velocity, $\vec{B}$ denotes the magnetic field strength, $\rho$ denotes the plasma density and $v_A$ denotes the background Alfv\'en speed. $\vec{B}_0$ and $\rho_0$ denote the equilibrium magnetic field and density and $\vec{u}$, $\vec{b}$, $\rho_1$ denote the perturbation components. We justify assumption \eqref{eq:linearisation_assumption1} because observational evidence suggests that $u / v_A$ lies approximately in the range $[10^{-2},10^{-1}]$ \citep{McIntosh2011,McIntosh2012}. We approximate $\pdv*{B_0}{t}$ by zero, since magnetic structures can last much longer than the typical Alfv\'en travel time of waves in coronal loops. If we take the Alfv\'en speed in the corona as about $1\si{.Mm.s^{-1}}$ and assume that coronal loops have a characteristic length of about $100\si{.Mm}$ then this gives an Alfv\'en travel time of about $100\si{.s}$. 
Magnetic structures can last much longer than this, for example, active region prominences can last a few hours or a day, while quiescent prominences can last between a few and 300 days \citep{Priest2014}. We assume that $\pdv*{\rho_0}{t}$ approximately holds in coronal loops which are isothermal. \citet{Klimchuk2015} shows that loops can be approximated as isothermal if the time interval between heating events/nanoflares in the loops is sufficiently small.

We simplify further by modelling the background magnetic field strength as uniform and given by
\begin{equation}
    \vec{B}_0 = B_0 \vec{\hat{B}}_0.
\end{equation}
We define the unit vectors $\vec{\hat{B}}_0$, $\vec{\hat{\perp}}$ as
\begin{eqnarray}
    \vec{\hat{B}}_0 & = & \sin\alpha\, \vec{\hat{y}} + \cos\alpha\, \vec{\hat{z}}, \\
    \vec{\hat{\perp}} & = & \cos\alpha\, \vec{\hat{y}} - \sin\alpha\, \vec{\hat{z}}.
\end{eqnarray}
as the unit vectors parallel and perpendicular to the equilibrium magnetic field.

The background Alfv\'en speed is given by $v_A(x,z) = \hat{v}_A(x)v_A(0,z)$, where
\begin{equation}
    \label{eq:alfven_speed_along_z}
    v_A(0,z) = \begin{cases}
    v_{A-}, & z < 0, \\
    v_{A+}, & z \ge 0,
    \end{cases}
\end{equation}
and $\hat{v}_A(x)$ is an arbitrary function of $x$ satisfying $\hat{v}_A(0)=1$. Note that we have neglected the exponential stratification due to gravity for mathematical convenience.

We model the plasma as cold and ideal. Therefore, the linearised momentum equation is given by
\begin{equation}
    \label{eq:vector_momentum_eqn}
    \pdv{\vec{u}}{t} = \frac{1}{\mu\rho_0(x,z)}\qty[(\vec{B}_0\vdot\grad)\vec{b}-\grad(\vec{b}\vdot\vec{B}_0)],
\end{equation}
and the induction equation by
\begin{equation}
    \label{eq:vector_induction_eqn}
    \pdv{\vec{b}}{t}=(\vec{B}_0\vdot\grad)\vec{u} - \vec{B}_0\div{\vec{u}},
\end{equation}
where
\begin{equation}
    v_A(x,z) = \frac{B_0}{\sqrt{\mu \rho(x,z)}}.
\end{equation}
Since we set the plasma beta equal to zero, the velocity perturbations parallel to the background magnetic field decouple from the velocity perturbations perpendicular to the field. 
Hence, the velocity is given by
\begin{equation}
    \vec{u} = u_x \vec{\hat{x}} + u_\perp \vec{\hat{\perp}}
\end{equation}
and the magnetic field is given by
\begin{equation}
\begin{aligned}
    \vec{\hat{b}} &= \vec{b} / B_0 \\
    &= \hat{b}_x \vec{\hat{x}} + \hat{b}_\perp \vec{\hat{\perp}} + \hat{b}_{||} \vec{\hat{B}}_0\; .
\end{aligned}
\end{equation}
Hence, Equations \eqref{eq:vector_momentum_eqn} and \eqref{eq:vector_induction_eqn} simplify to give
\begin{eqnarray}
    \label{eq:ux1}
    \pdv{u_x}{t} & = & v_A^2(x,z)\qty[\nabla_{||}\hat{b}_x - \pdv{\hat{b}_{||}}{x}], \\
    \label{eq:u_perp1}
    \pdv{u_\perp}{t} & = & v_A^2(x,z)\Big[\nabla_{||}\hat{b}_\perp - \nabla_\perp \hat{b}_{||} \Big], \\
    \label{eq:bx1}
    \pdv{\hat{b}_x}{t} & = & \nabla_{||}u_x, \\
    \label{eq:b_perp1}
    \pdv{\hat{b}_\perp}{t} & = & \nabla_{||} u_\perp, \\
    \label{eq:b_par1}
    \pdv{\hat{b}_{||}}{t} & = & -\qty[\pdv{u_x}{x} + \nabla_\perp u_\perp],
\end{eqnarray}
where
\begin{eqnarray}
    \nabla_\perp & = & \cos\alpha \pdv{}{y} - \sin\alpha \pdv{}{z}, \\
    \nabla_{||} & = & \sin\alpha\pdv{}{y} + \cos\alpha\pdv{}{z},
\end{eqnarray}
are the derivatives perpendicular and parallel to the
equilibrium magnetic field respectively.
We can eliminate $\hat{b}_x$, $\hat{b}_\perp$, $\hat{b}_{||}$ to give
\begin{eqnarray}
    \label{eq:ux_in_terms_of_u_perp}
    \qty[\pdv[2]{}{t} - v_A^2\qty(\pdv[2]{}{x}+\nabla_{||}^2)]u_x & = & v_A^2\nabla_\perp\pdv{u_\perp}{x}, \\
    \qty[\pdv[2]{}{t} - v_A^2\qty(\nabla_\perp^2 + \nabla_{||}^2)]u_\perp  & = & v_A^2\nabla_\perp \pdv{u_x}{x}.
\end{eqnarray}

In this paper, we will derive exact analytic solutions to the equations. However, the formulas we derive are often cumbersome, and so our discussion will focus on approximate (but less cumbersome) solutions in our narrower parameter space. The parameter space we will focus on is  
\begin{eqnarray}
    \label{eq:parm_space_1}
    k_x \gg k_{||+}, \\
     \label{eq:parm_space_2}
    k_{||-} \gg k_{||+}, \\
    \label{eq:parm_space_ky}
    0 < k_y \le O(k_{||+}), \\
    \label{eq:parm_space_4}
    0 < \tan\alpha \le O(1), 
\end{eqnarray}
where 
\begin{equation}
    k_{||\pm} = \frac{\omega}{v_{A\pm}},
\end{equation}
gives the Alfv\'en wavenumber and $\omega$ gives the wave frequency. We model $k_x\gg k_{||+}$ as this approximates the conditions near a singularity in a resonant absorption experiment. The results should also apply to other cases where $k_x\gg k_{||+}$, for example, during the nonlinear development of an instability. Our condition for $\alpha$ prevents the background magnetic field from being tangential to the $z=0$ plane.

\section{Uniform background Alfv\'en speed}
\label{sec:uniform_alfven_speed}

In this section, we will calculate the solution in a domain where the background Alfv\'en speed is uniform, i.e. $v_A(x,z)=v_{A+}$ and we impose an incident Alfv\'en wave from $z>0$. We will impose line-tied boundary conditions at $z=0$, i.e. we will ensure that $u_x=u_\perp=0$ at $z=0$. The model we use here is very similar to that used in Section 2 of \citet{Halberstadt1993} and Section 5 of \citet{Goedbloed1994}. The only significant difference is that they use a closed system with line-tied boundary conditions imposed at both ends in $z$ whereas we model an open system with line-tied boundary conditions at $z=0$. We find that modelling an open-loop makes the maths a little bit easier. \citet{Goedbloed1994} shows that imposing line-tied boundary conditions forces the fast and Alfv\'en waves to couple due to the impossibility to reconcile the requirements of the vanishing of both $u_x$ and $u_\perp$. Therefore, pure Alfv\'en wave solutions and pure fast wave solutions cannot exist. The fast waves can be evanescent and form boundary layers at $z=0$. Our goal in this section is to reproduce these boundary layers to explain how they form and motivate why testing the validity of line-tied boundary conditions is necessary.

By assuming the variables are of the form 
\begin{equation}
    f(\vec{x},t) = f_0 \exp[i(k_x x + k_y y + k_{zn} + \omega t)],
\end{equation}
where $f$ can be $u_x$, $u_\perp$, $b_x$, $b_\perp$ or $b_{||}$, we can derive the following dispersion relation
\begin{equation}
    \label{eq:dispersion_relation_uniform}
    \Big[\omega^2 + v_{A+}^2\nabla_{||n}^2\Big]\Big[\omega^2 - v_{A+}^2(k_x^2 + k_y^2 + k_{zn}^2)\Big]=0,
\end{equation}
where 
\begin{equation}
    \nabla_{||n} = i(k_y \sin\alpha + k_{zn}\cos\alpha),
\end{equation}
and $n$ is an integer used to distinguish the different solutions.
Solutions to Equation \eqref{eq:dispersion_relation_uniform}, where the left bracket equals zero, correspond to Alfv\'en waves that propagate parallel to the background magnetic field. Solutions, where the right bracket equals zero, correspond to fast waves, and these waves propagate isotropically.
Hence, the possible values of $k_{zn}$ are
\begin{equation}
    \begin{pmatrix}
    k_{z1} \\
    k_{z2} \\
    k_{z3} \\
    k_{z4}
    \end{pmatrix}
    =
    \begin{pmatrix}
    k_{||+} / \cos\alpha - k_y \tan\alpha \\
    -k_{||+} / \cos\alpha - k_y \tan\alpha \\
    i\sqrt{k_x^2 + k_y^2 - k_{||+}^2} \\
    -i\sqrt{k_x^2 + k_y^2 - k_{||+}^2}
    \end{pmatrix}\; ,
\end{equation}
for the upward and downward propagating Alfv\'en waves and the
upward and downward propagating fast waves. Note that, as $k_x/k_{||+} \rightarrow \infty$, 
the fast waves are evanescent ($k_{z3}$) and 
exponentially growing ($k_{z4}$), instead of propagating.

We assume the solutions are of the form
\begin{equation}
    f(\vec{x},t) = \sum_{n=1}^4 f_n \exp[i(k_x x + k_y y + k_{zn} z + \omega t)].
\end{equation}
We impose an incident Alfv\'en wave from above ($z>0$) by setting $u_{\perp 1}=u_0$, where $u_0$ gives the incident wave amplitude and set the amplitude of the incident fast waves to zero,
$f_4=0$. From Equation \eqref{eq:ux_in_terms_of_u_perp} we know that
\begin{equation}
    \label{eq:ux_u_perp}
    u_{xn} = \hat{u}_{xn} u_{\perp n},
\end{equation}
where
\begin{eqnarray}
    \label{eq:uxn}
    \hat{u}_{xn} & = & \frac{-ik_x\nabla_{\perp n}}{\mathcal{L}_{n} - k_x^2}, \\
    \mathcal{L}_{n} & = & \nabla_{||n}^2 + k_{||+}^2,\\
    \nabla_{\perp n} & = & i(k_y \cos\alpha - k_{zn}\sin\alpha).
\end{eqnarray}
The boundary conditions require $u_x=u_\perp=0$ at $z=0$ and this gives the following matrix equation
\begin{equation}
    \begin{pmatrix}
    1 & 1 \\
    \hat{u}_{x2} & \hat{u}_{x3}
    \end{pmatrix}
    \begin{pmatrix}
    u_{\perp2} \\
    u_{\perp3}
    \end{pmatrix}=-u_0
    \begin{pmatrix}
    1 \\
    \hat{u}_{x1}
    \end{pmatrix}.
\end{equation}
Hence,
\begin{equation}
\begin{pmatrix}
u_{\perp1} \\
u_{\perp2} \\
u_{\perp3}
\end{pmatrix}=u_0
\begin{pmatrix}
1 \\
-(\hat{u}_{x1} - \hat{u}_{x3})/(\hat{u}_{x2} - \hat{u}_{x3}) \\
(\hat{u}_{x1} - \hat{u}_{x2})/(\hat{u}_{x2} - \hat{u}_{x3}) \\
\end{pmatrix}.
\end{equation}
From Equations \eqref{eq:bx1}, \eqref{eq:b_perp1} and \eqref{eq:b_par1}, we know that
\begin{eqnarray}
    \label{eq:bx_u_perp}
    v_{A+}\hat{b}_{xn} & = & \frac{\nabla_{||n}}{ik_{||+}}u_{xn}, \\
    \label{eq:b_perp_u_perp}
    v_{A+}\hat{b}_{\perp n} & = & \frac{\nabla_{||n}}{ik_{||+}}u_{\perp n}, \\
    \label{eq:b_par_u_perp}
    v_{A+}\hat{b}_{|| n} & = & -\frac{1}{ik_{||+}}\qty[ik_x u_{xn}+\nabla_{\perp n} u_{\perp n}].
\end{eqnarray}

This gives the full set of solutions to the equations. However, we do not write out the solutions in full here as they are cumbersome. We find the equations are easier to interpret if we construct series solutions by taking appropriate limits. Resonant absorption cannot occur because the background Alfv\'en speed is uniform. Therefore, singularities cannot form in our model. To simulate the conditions near a singularity in a resonant absorption experiment where $\pdv*{v_A}{x}\ne0$, 
we calculate the asymptotic expansions for $k_x/k_{||+}\rightarrow\infty$.
The leading order asymptotic expansions for the Alfv\'en wave coefficients is given by
\begin{eqnarray}
    \label{eq:uniform_alfven_wave_asymptotic_expansion_1}
    \hat{b}_{||1} = \hat{b}_{||2} & = & 0, \\
    \frac{u_{x1}}{u_0},\ \frac{u_{x2}}{u_0},\ \frac{v_{A+}\hat{b}_{x1}}{u_0},\ \frac{v_{A+}\hat{b}_{x2}}{u_0} & = & O\qty(\frac{k_{||+}}{k_x}), \\
    \frac{u_{\perp1}}{u_0},\ \frac{u_{\perp 2}}{u_0},\ \frac{v_{A+}\hat{b}_{\perp1}}{u_0},\ \frac{v_{A+}\hat{b}_{\perp2}}{u_0} & = & O(1),
\end{eqnarray}
where the full leading order expressions are given in Appendix \ref{adx:uniform_background_alfven_speed}.
For large $k_x/k_{||+}$, the fast wave coefficients are given by
\begin{eqnarray}
    \frac{u_{x3}}{u_0},\ \frac{u_{\perp 3}}{u_0} & = & O\qty(\frac{k_{||+}}{k_x}), \\
    \label{eq:uniform_fast_wave_asymptotic_expansion_2}
    \frac{v_{A+}\hat{b}_{x3}}{u_0},\ \frac{v_{A+}\hat{b}_{\perp3}}{u_0},\ \frac{v_{A+}\hat{b}_{||3}}{u_0} & = &O(1).
\end{eqnarray}
This shows that the leading order solution for $u_x$, $b_x$, $b_{\perp}$ and $b_{||}$ contain evanescent fast wave terms and this gives rise to the boundary layers seen in, for example,  \citet{Halberstadt1993,Halberstadt1995,Arregui2003}.

Now, we give a brief explanation as to why the boundary layers form. Imposing $\vec{u}=0$ at $z=0$ forces the normal component of the magnetic field, $b_z$, to be zero. For proof, consider the $z$-component of the induction equation,
\[\pdv{b_z}{t}=\vec{\hat{z}}\cdot\curl(\vec{u}\cross\vec{B}_0).\]
The curl in the $z$-direction only depends on derivatives in $x$ and $y$. Since  $\vec{u}=0=$ constant at $z=0$, the $x$ and $y$ derivatives are zero. Therefore, if $b_z$ is initially zero, it will remain zero for all time. Therefore, at $z=0$,
\[b_z = \cos\alpha b_{||} - \sin\alpha b_\perp=0,\]
rearranging gives
\[b_{||} = \tan\alpha\, b_\perp.\]
Here, $b_\perp$ gives the magnetic field component of the Alfv\'en wave and $b_{||}$ gives the magnetic pressure component of the fast wave. Hence, for $\alpha\ne0$, if there is a large-amplitude Alfv\'en wave, then there must also be a large-amplitude fast wave (for $\alpha \ne 0$). If $v_A^2(k_x^2+k_y^2)>\omega^2$, this forces the associated $k_z$ to be imaginary, which means the fast wave must be evanescent and so boundary layers will form.

Note that for $k_x\rightarrow\infty$, $k_{z3}\rightarrow ik_x$. Therefore, the time-averaged energy associated with the evanescent fast waves is proportional to $\exp(-2k_x z)$ and the total fast wave
energy is proportional to
\begin{equation}
    k_{||+}\int_0^\infty \exp(-2 k_x z) dz = \frac{k_{||+}}{2k_x}.
\end{equation}
Thus, the evanescent fast waves' energy tends to 0 as $k_x\rightarrow \infty$, for $u_0$ fixed. Hence, the evanescent fast waves will have a limited effect on the rate of resonant absorption. Since the energy associated with the boundary layers does not grow as the length scales in $x$ get shorter, they cannot be absorbing energy. Moreover, the boundary layers cannot directly transport energy away as they are evanescent.

\section{Piecewise constant background Alfv\'en speed}
\label{sec:piecewise_constant_alfven_speed}

In this section we expand upon the previous section by keeping the Alfv\'en speed uniform in $x$ but using a piecewise constant profile in $z$ instead, i.e. the Alfv\'en speed is given by
\begin{equation}
    v_A(z) = \begin{cases}
    v_{A-}, & z < 0 \\
    v_{A+}, & z \ge 0.
    \end{cases}
\end{equation}
The region $z>0$ models the corona and $z<0$ models the chromosphere, therefore, $v_{A+} > v_{A-}$. We impose an incident Alfv\'en wave from the region $z>0$ and then calculate the unique solution which ensures continuity of $u_x$, $u_\perp$, $\pdv*{u_x}{z}$, $\pdv*{u_\perp}{z}$ at $z=0$. By checking if our solutions agree with the previous section, we can test the validity of imposing line-tied boundary conditions.

We assume a solution of the form
\begin{equation}
    f = \sum_{n=1}^4 f_{n+}\exp[i(k_x x + k_y y + k_{zn+} z + \omega t)],
\end{equation}
for $z\ge 0$ and
\begin{equation}
    f = \sum_{n=1}^4 f_{n-}\exp[i(k_x x + k_y y + k_{zn-} z + \omega t)],
\end{equation}
for $z<0$ where
\begin{equation}
\begin{pmatrix}
k_{z1\pm} \\
k_{z2\pm} \\
k_{z3\pm} \\
k_{z4\pm} \\
\end{pmatrix}
=
\begin{pmatrix}
 k_{||\pm}/\cos\alpha - k_y\tan\alpha \\
-k_{||\pm}/\cos\alpha - k_y\tan\alpha \\
 i\sqrt{k_x^2+k_y^2 - k_{||\pm}^2} \\
-i\sqrt{k_x^2+k_y^2 - k_{||\pm}^2}
\end{pmatrix}.
\end{equation}
We impose an incident Alfv\'en wave from $z>0$ by setting $u_{\perp1+}=u_0$, where $u_0$ gives the incident wave amplitude. We impose the amplitude of the incident fast waves to be zero. Therefore, $f_{4+}=0$. There are no incident waves from $z<0$, hence, $f_{2-}=f_{3-}=0$.  From Equation \eqref{eq:ux_in_terms_of_u_perp} we know that
\begin{equation}
    \label{eq:ux_pm_u_perp_pm}
    u_{x n\pm} = \hat{u}_{x n\pm}u_{\perp n\pm},
\end{equation}
where
\begin{eqnarray}
    \label{eq:uxnpm}
    \hat{u}_{x n\pm} & = & \frac{-ik_x\nabla_{\perp n\pm}}{\mathcal{L}_{n\pm}-k_x^2}, \\
    \mathcal{L}_{n\pm} & = &\nabla_{||n\pm}^2+k_{||\pm}^2, \\
    \nabla_{\perp n\pm} & = & i(k_y\cos\alpha - k_{zn\pm}\sin\alpha), \\
    \nabla_{||n\pm} & = & i(k_y \sin\alpha + k_{zn\pm}\cos\alpha).
\end{eqnarray}
We require continuity of $u_x$, $\pdv*{u_x}{z}$, $u_\perp$ and $\pdv*{u_\perp}{z}$ at $z=0$. This gives the following matrix equation
\begin{equation}
    \begin{pmatrix}
    \hat{u}_{x1-} & \hat{u}_{x4-} & -\hat{u}_{x2+} & -\hat{u}_{x3+} \\
    k_{z1-}\hat{u}_{x1-} & k_{z4-}\hat{u}_{x4-} & -k_{z2+}\hat{u}_{x2+} & -k_{z3+}\hat{u}_{x3+} \\
    1 & 1 & -1 &-1 \\
    k_{z1-} & k_{z4-} & -k_{z2+} & -k_{z3+}
    \end{pmatrix}
    \begin{pmatrix}
    u_{\perp 1-} \\
    u_{\perp 4-} \\
    u_{\perp 2+} \\
    u_{\perp 3+}
    \end{pmatrix}
    =u_0
    \begin{pmatrix}
    \hat{u}_{x1+} \\
    k_{z1+}\hat{u}_{x1+} \\
    1 \\
    k_{z1+}
    \end{pmatrix}.
\end{equation}
From Equations \eqref{eq:bx1}, \eqref{eq:b_perp1} and \eqref{eq:b_par1}, we know that
\begin{eqnarray}
    \label{eq:bx_pm_u_perp_pm}
    v_{A+}\hat{b}_{xn\pm}  & = & \frac{\nabla_{||n\pm}}{ik_{||+}}u_{xn\pm}, \\
    \label{eq:b_perp_pm_u_perp_pm}
    v_{A+}\hat{b}_{\perp n\pm} & = & \frac{\nabla_{||n\pm}}{ik_{||+}}u_{\perp n\pm}, \\
    \label{eq:b_par_pm_u_perp_pm}
    v_{A+}\hat{b}_{|| n\pm} & = & -\frac{1}{ik_{||+}}\qty[ik_x u_{xn\pm}+\nabla_{\perp n\pm} u_{\perp n\pm}].
\end{eqnarray}

\begin{figure*}
    \centering
    \includegraphics[width=\textwidth,height=0.91\textheight,keepaspectratio]{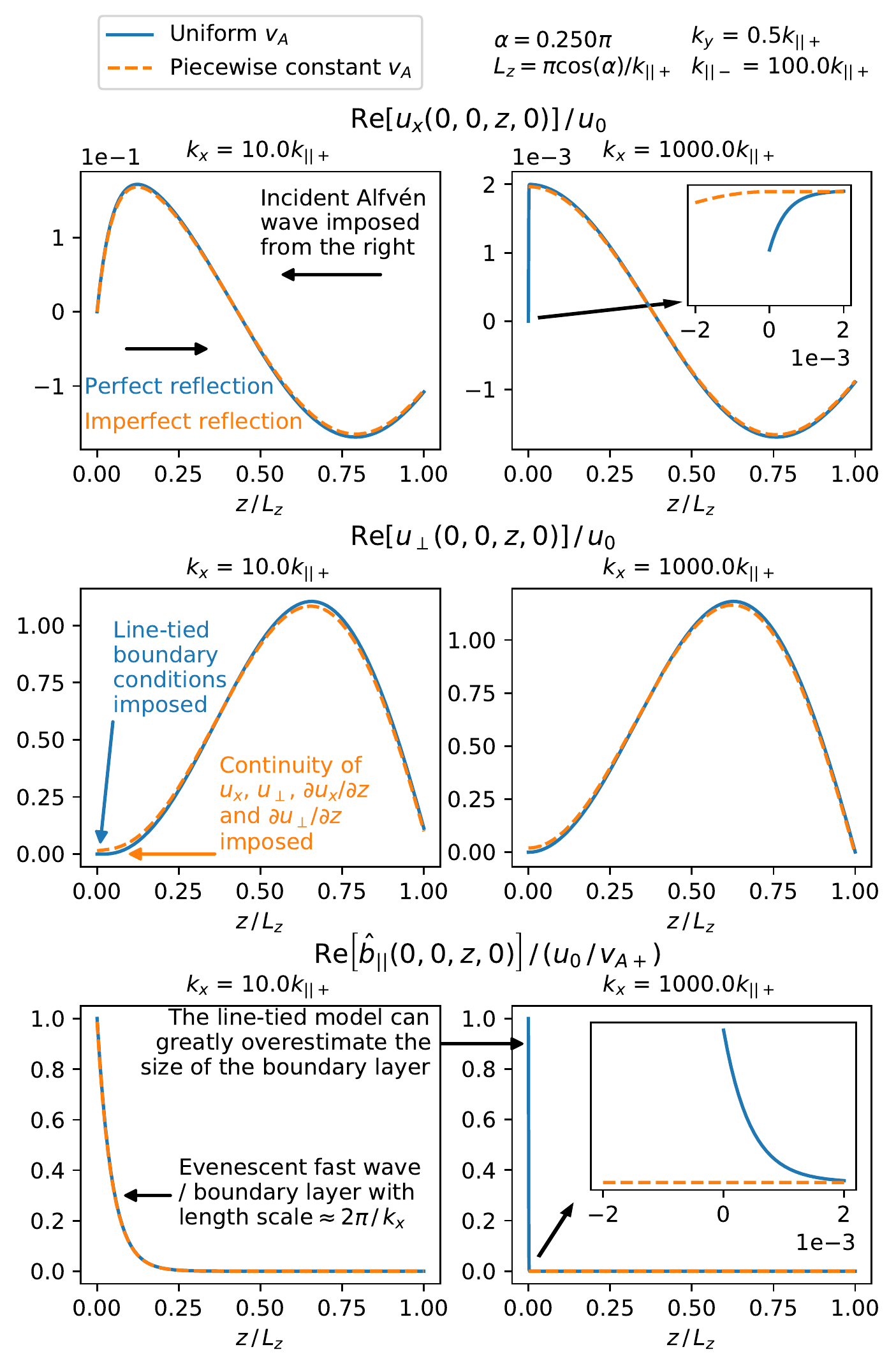}
    \caption{This figure shows plots of the real part of $u_x$ (top row), $u_\perp$ (middle row) and $\hat{b}_{||}$ (bottom row) at $x=y=t=0$ as a function of $z$ for $k_x=10k_{||+}$ (left column) and $k_x = 1000k_{||+}$ (right column). The case where the background Alfv\'en speed is uniform and line-tied boundary conditions is imposed is shown in blue and the case where the Alfv\'en speed is piecewise constant and continuity of $\vec{u}$ is imposed is shown in dashed orange. A description of this figure is continued in the caption of Figure \ref{fig:piecewise_constant_vs_uniform_imag_part}.}
    \label{fig:piecewise_constant_vs_uniform_real_part}
\end{figure*}

\begin{figure*}
    \centering
    \includegraphics[width=\textwidth,height=0.88\textheight,keepaspectratio]{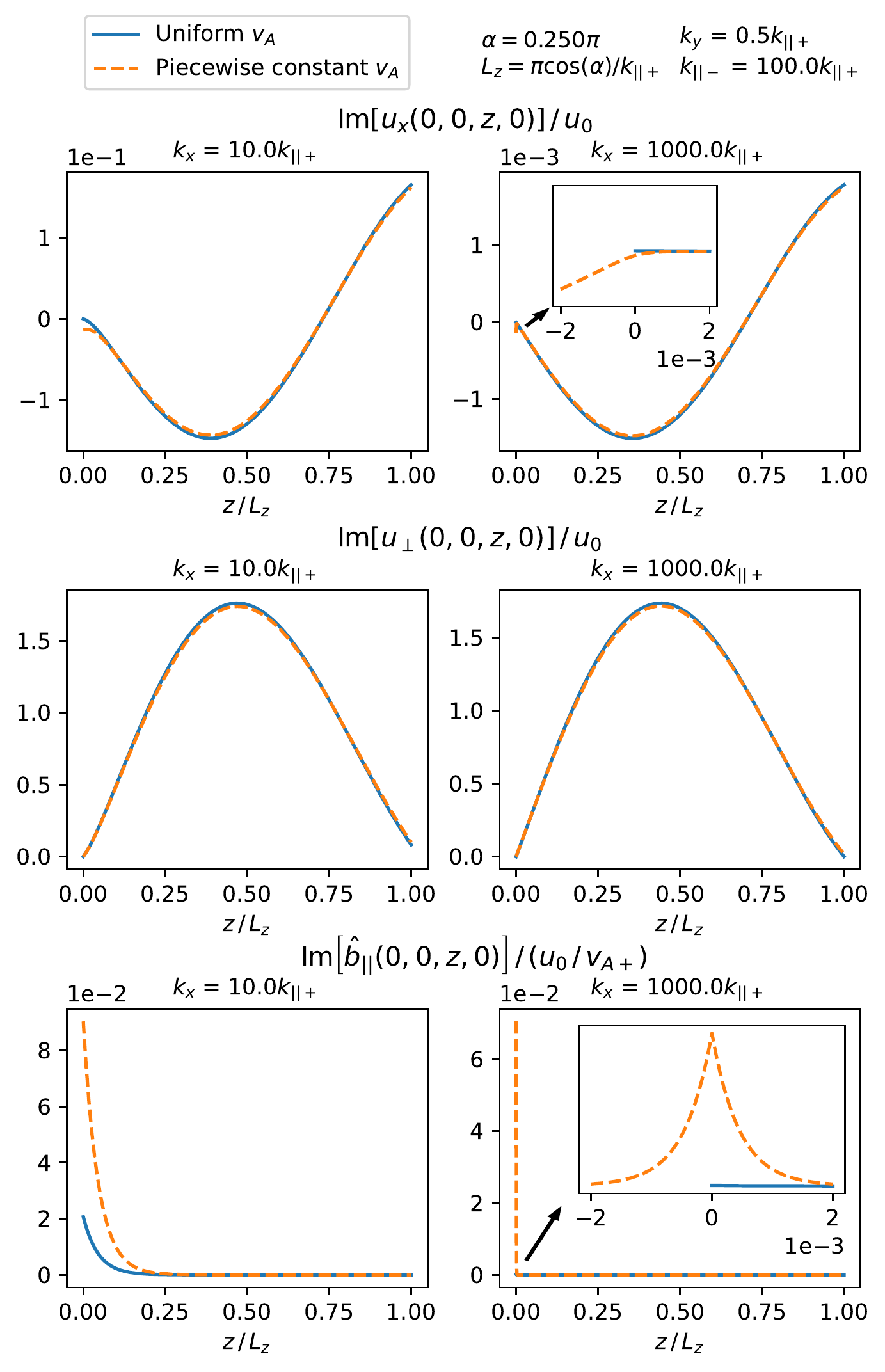}
    \caption{This figure is similar to Figure \ref{fig:piecewise_constant_vs_uniform_real_part} except it plots the imaginary part of each variable instead. The blue and orange curves show good agreement in the left-hand column for both figures. The small differences occur due to the blue curves' waves undergoing perfect reflection and the orange waves undergoing imperfect reflection at $z=0$. The right-hand column shows less agreement between the plots, and this is because the fast waves are evanescent, whereas in the left-column the fast waves can propagate for $z<0$. Equations  \eqref{eq:uniform_alfven_wave_asymptotic_expansion_1}-\eqref{eq:uniform_fast_wave_asymptotic_expansion_2} and \eqref{eq:piecewise_alfven_wave_asymptotic_expansion_1}-\eqref{eq:piecewise_fast_wave_asymptotic_expansion_2} show that the fast wave terms in the blue solutions are of a higher order than the orange solutions.}
    \label{fig:piecewise_constant_vs_uniform_imag_part}
\end{figure*}

This gives the exact solution to the equations. However, the solution is too cumbersome to present directly. To build our intuition for how the solutions change with $k_x$ we plot the real/imaginary part of $u_x$, $u_\perp$ and $\hat{b}_{||}$  as a function of $z$ at $x=y=t=0$ in Figure \ref{fig:piecewise_constant_vs_uniform_real_part}/\ref{fig:piecewise_constant_vs_uniform_imag_part} respectively. We show the parameters we use in the top-right of each figure. The blue curve shows the solutions from Section \ref{sec:uniform_alfven_speed} where line-tied boundary conditions ($\vec{u}=0$) are imposed at $z=0$. The dashed orange curve shows the solution with continuity of $\vec{u}$ and $\pdv*{\vec{u}}{z}$ imposed at $z=0$ with a piecewise constant background Alfv\'en speed. In the left-hand column, $k_x=10k_{||+}$, this means that the fast waves are evanescent for $z>0$ and can propagate for $z<0$. It shows that the solutions are in good agreement and suggests that line-tied boundary conditions give a good approximation for the solution where $z>0$. The small disagreement occurs due to line-tied boundary conditions causing perfect reflection, whereas, a finite jump in the background Alfv\'en speed results in partial reflection. In the right column, $k_x=1000k_{||+}$, this means that the fast waves are evanescent everywhere. The plots show poor agreement near $z=0$ and good agreement for $z\gg0$. They show that the model with line-tied boundary conditions can significantly overestimate the evanescent fast waves'/boundary layers' amplitude.

\begin{figure*}
    \centering
    \includegraphics[width=17cm,height=0.88\textheight,keepaspectratio]{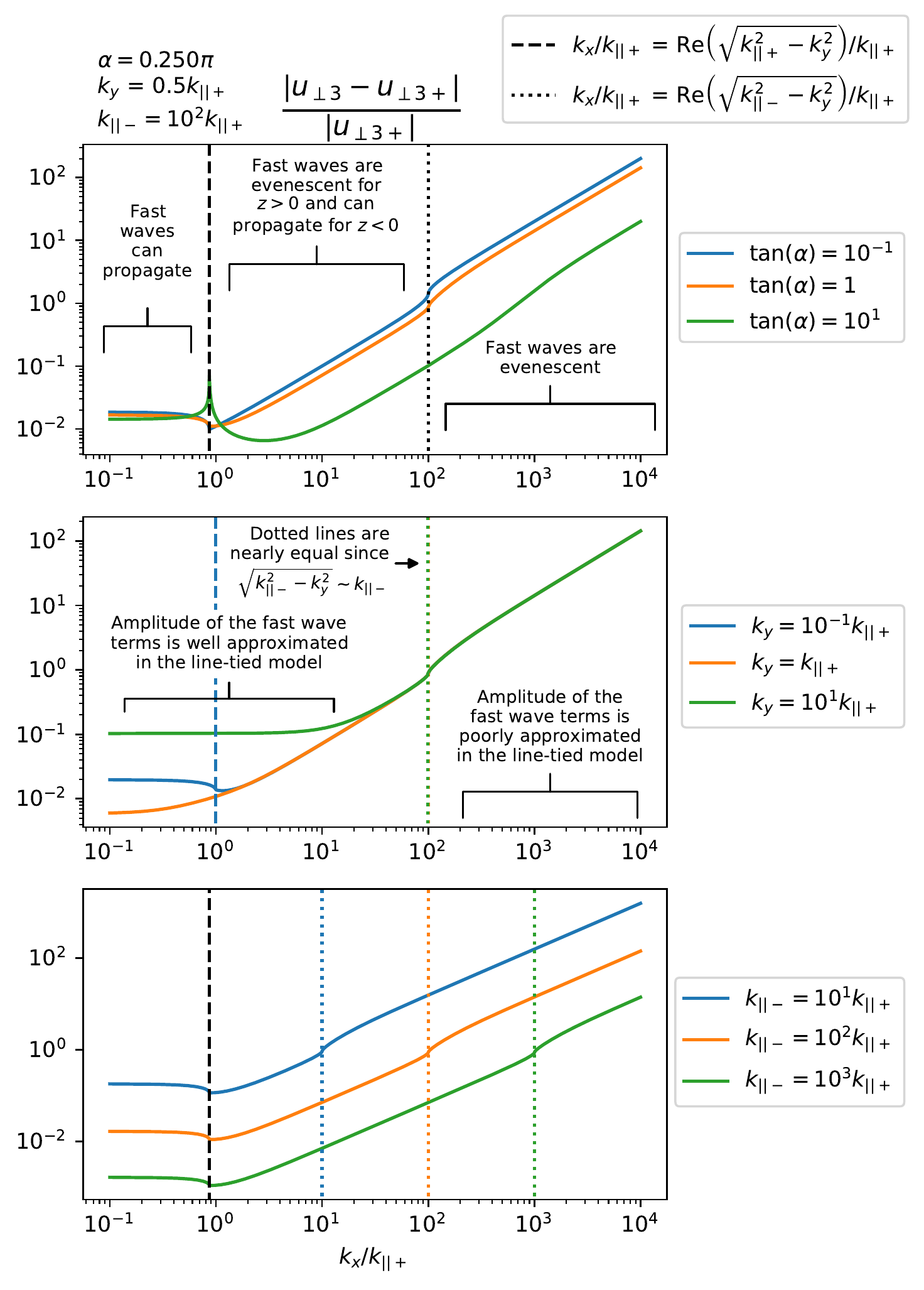}
    \caption{This figure plots the relative difference (see Equation \ref{eq:relative_difference}) introduced by using line-tied boundary conditions to estimate the size of the boundary layers as a function of $k_x$. Unless stated otherwise, we use the parameters shown in the top-left. In the top plot, the different colours correspond to different values of $\alpha$, in the middle plot they correspond to different values of $k_y$ and in the bottom plot they correspond to different values of $k_{||-}$. If $k_x$ is to the left/right of the vertical dashed lines, then the fast waves can propagate/are evanescent for $z>0$ respectively. Similarly, if $k_x$ is to the left/right of the vertical dotted lines, the fast waves can propagate/are evanescent for $z<0$ respectively.}
    \label{fig:fast_wave_error_vs_kx}
\end{figure*}

We have seen that depending on the value of $k_x$, our models can sometimes agree on the evanescent fast waves' amplitude and sometimes disagree strongly. Our goal now is to quantify how well the model with line-tied boundary conditions approximates the amplitude of the evanescent fast wave/boundary layer as a function of $k_x$, for the parameter space given by Equations \eqref{eq:parm_space_1}-\eqref{eq:parm_space_4}. Figure \ref{fig:fast_wave_error_vs_kx} plots the relative difference, given by,
\begin{equation}
    \label{eq:relative_difference}
    \text{relative difference} = \abs{\frac{u_{\perp3} - u_{\perp 3+}}{u_{\perp 3+}}},
\end{equation}
where $u_{\perp3}$ gives the amplitude of the fast wave component of $u_\perp$ at $z=0$ in the model where line-tied boundary conditions are imposed and $u_{\perp 3+}$ gives the amplitude in the model where a piecewise constant background Alfv\'en speed is used. Equations \eqref{eq:ux_u_perp}, \eqref{eq:bx_u_perp}, \eqref{eq:b_perp_u_perp}, \eqref{eq:b_par_u_perp} show that $u_{xn}$, $b_{xn}$, $b_{\perp n}$, $b_{|| n}$ are equal to scalar multiples of $u_{\perp n}$, similarly, Equations \eqref{eq:ux_pm_u_perp_pm}, \eqref{eq:bx_pm_u_perp_pm}, \eqref{eq:b_perp_pm_u_perp_pm}, \eqref{eq:b_par_pm_u_perp_pm} show that $u_{xn+}$, $b_{xn+}$, $b_{\perp n+}$, $b_{|| n+}$ are equal to the same scalar multiples of $u_{\perp n+}$ respectively. Therefore,
\[
    \abs{\frac{u_{\perp3} - u_{\perp 3+}}{u_{\perp 3+}}} = \abs{\frac{u_{x3} - u_{x 3+}}{u_{x 3+}}} = ... = \abs{\frac{b_{||3} - b_{|| 3+}}{b_{|| 3+}}}.
\]
We show all the plots as functions of $k_x$, and the parameters take values shown in the top-left (unless stated otherwise). The vertical dashed lines show where 
\[k_x = \Re\qty(\sqrt{k_{||+}^2-k_y^2})\] 
and the dotted lines show where 
\[k_x = \Re\qty(\sqrt{k_{||-}^2-k_y^2})= k_{||-} + O\qty(\frac{k_{||+}}{k_{||-}}),\]
where we assume $k_y$ satisfies Equation \eqref{eq:parm_space_ky}.
If $k_x$ is to the right of these vertical lines, then the fast waves are evanescent, and if $k_x$ is to the left, then the fast waves can propagate. The plots show that the relative difference between the models is dependent on whether the fast waves are evanescent or propagating. If $k_x \gg k_{||-}$, then the models show poor agreement and the model with line-tied boundary conditions overestimates the size of the boundary layers. On the other hand, if $k_x \ll k_{||-}$
then the models show good agreement.

Since $k_{||-}\gg k_{||+}$ it is not clear if $k_x \gg k_{||-}$, even if the waves undergo resonant absorption. Our goal now is to check if it is possible for $k_x\gg k_{||-}$ in the corona by calculating the steady-state wavenumber, $k_x^*$, for driven standing phase-mixed Alfv\'en waves in a viscous and resistive domain (note that waves during resonant absorption undergo a similar phase-mixing process). \citet{Heyvaerts1983,Priest2014} calculate this value as
\begin{equation}
\begin{aligned}
    \label{eq:kx_star}
    k_x^* &= \qty[\frac{6\omega}{(\eta+\nu)a}]^{1/3} \\
    &\approx 1.82 \times10^{-3} \qty[\frac{\omega}{10^{-2}\si{.s^{-1}}}\frac{10\si{.m^2.s^{-1}}}{\eta+\nu}\frac{10^6\si{.m}}{a}]^{1/3}\si{.m^{-1}}
\end{aligned}
\end{equation}
for typical coronal values, where $\eta+\nu$ denotes the sum of the magnetic diffusivity and kinematic viscosity coefficients and $a$ is given by
\begin{equation}
    a = \frac{v_A}{\pdv*{v_A}{x}}.
\end{equation}
The Alfv\'en wavenumber parallel to the background magnetic field in the chromosphere, $k_{||-}$, is given by
\small
\begin{equation}
    \label{eq:k_par-_typical_values}
    k_{||-} = 10^{-5}\qty(\frac{\omega}{10^{-2}\si{.s^{-1}}})\qty(\frac{10^3\si{.m.s^{-1}}}{v_{A-}})\si{.m^{-1}}.
\end{equation}
\normalsize
for typical values. Note that our value for $k_x^*$ represents an upper bound for $k_x$ in the corona. Waves may not be able to phase-mix to such short-length scales due to them leaking out of the corona \citep{Prokopyszyn2019} or because of the plasma's thermodynamic response \citep{Cargill2016}. Comparing Equations \eqref{eq:kx_star} and \eqref{eq:k_par-_typical_values}, they suggest that $k_x^*$ can greatly exceed $k_{||-}$ for waves undergoing resonant absorption in the corona. However, it takes time for waves to phase-mix to such short length scales. Therefore, for early times $k_x\ll k_{||-}$ is valid and after the waves have sufficiently phase-mixed $k_x \gg k_{||+}$ is valid. Our line-tied boundary condition solutions from Section \ref{sec:uniform_alfven_speed} provide a good approximation for the case where $k_x \ll k_{||-}$ as imposing line-tied boundary conditions is equivalent to modelling the chromospehre as infinitely denser than the corona which causes $k_{||-}\rightarrow \infty$. In Appendix \ref{adx:check_recover_of_uniform_equations} we check that the asymptotic expansions from Section \ref{sec:uniform_alfven_speed} are recovered if $k_{||+} \ll k_x\ll k_{||-}$.

In Appendix \ref{adx:piecewise_constant_background_alfven_speed} we calculate the solution for the case where $k_x \gg k_{||-}, k_{||+}$,
as this simulates the conditions for waves undergoing resonant absorption after they have phase-mixed to sufficiently short length scales. They show that the leading order asymptotic expansions for the Alfv\'en wave coefficients are given by
\begin{eqnarray}
    \label{eq:piecewise_alfven_wave_asymptotic_expansion_1}
    \hat{b}_{||1-} = \hat{b}_{||1+} = \hat{b}_{||2+} & = & 0, \\
    \frac{u_{x1-}}{u_0},\ \frac{u_{x1+}}{u_0},\ \frac{u_{x2+}}{u_0},\ \frac{v_{A+}\hat{b}_{x1-}}{u_0},\ \frac{v_{A+}\hat{b}_{x1+}}{u_0},\ \frac{v_{A+}\hat{b}_{x2+}}{u_0} & = & O\qty(\frac{k_{||-}}{k_x}), \\
    \frac{u_{\perp1-}}{u_0},\ \frac{u_{\perp1+}}{u_0},\ \frac{u_{\perp2+}}{u_0},\ \frac{v_{A+}\hat{b}_{\perp1-}}{u_0},\ \frac{v_{A+}\hat{b}_{\perp1+}}{u_0},\ \frac{v_{A+}\hat{b}_{\perp2+}}{u_0} & = & O(1),
\end{eqnarray}
where the full leading order expansions are given in Appendix \ref{adx:piecewise_constant_background_alfven_speed}.
The leading order asymptotic expansions for the coefficients of the fast wave are given by
\begin{eqnarray}
    \frac{u_{x4-}}{u_0},\ \frac{u_{x3+}}{u_0},\ \frac{u_{\perp4-}}{u_0},\ \frac{u_{\perp3+}}{u_0} & = & O\qty(\frac{k_{||-}^2}{k_x^2}), \\
    \label{eq:piecewise_fast_wave_asymptotic_expansion_2}
    \frac{v_{A+}\hat{b}_{x4-}}{u_0},\ \frac{v_{A+}\hat{b}_{x3+}}{u_0},\ \frac{v_{A+}\hat{b}_{\perp4-}}{u_0},\ \frac{v_{A+}\hat{b}_{\perp3+}}{u_0},\ \frac{v_{A+}\hat{b}_{||4-}}{u_0},\ \frac{v_{A+}\hat{b}_{||3+}}{u_0} & = & O\qty(\frac{k_{||-}}{k_x}), \\
\end{eqnarray}
Comparing Equations \eqref{eq:piecewise_alfven_wave_asymptotic_expansion_1}-\eqref{eq:piecewise_fast_wave_asymptotic_expansion_2} with Equations \eqref{eq:uniform_alfven_wave_asymptotic_expansion_1}-\eqref{eq:uniform_fast_wave_asymptotic_expansion_2}, we see that the Alfv\'en wave terms retain the same order and the fast wave terms are reduced here by a factor $k_x$. These results suggest that imposing line-tied boundary conditions can cause the model to significantly overestimate the amplitude of the evanescent fast waves / boundary layers for $\alpha\ne0$.
Equations \eqref{eq:piecewise_alfven_wave_asymptotic_expansion_1}-\eqref{eq:piecewise_fast_wave_asymptotic_expansion_2} show that the leading order velocity does not contain any fast-wave terms. Compare this with Equations \eqref{eq:uniform_alfven_wave_asymptotic_expansion_1}-\eqref{eq:uniform_fast_wave_asymptotic_expansion_2}, which show that $u_x$ does contain a fast wave component to leading order. Therefore, for a steady-state solution in a resonant absorption experiment, the fast wave components for $u_x$ and $u_\perp$ should be negligible compared with Alfv\'en wave components near the singularities, provided line-tied boundary conditions are not imposed. In the next section we verify this claim.

It is quite counter-intuitive that the length scales parallel to the chromosphere/corona interface, e.g. $k_x$, can have such a large effect on the validity of line-tied boundary conditions. Our goal now is to give an intuitive explanation for why this is. Imposing line-tied boundary conditions is in some sense equivalent to modelling the chromosome as infinitely denser than the corona, i.e. setting $k_{||-}\rightarrow \infty$. Therefore, imposing line-tied boundary conditions forces the fast waves in the chromosphere to be propagating waves. Propagating waves have a sinusoidal profile while evanescent waves have an exponential profile. This has important implications for ensuring continuity of velocity and its derivative at the chromosome/corona interface and explains why the line-tied solution can be so different to the solution with a piecewise constant density profile at the interface.

\section{Resonant absorption model}
\label{sec:resonant_absorption_model}

In this section, we expand on the previous section by allowing the Alfv\'en speed to depend on $x$. Therefore, the Alfv\'en speed is now given by $v_A(x,z)=\hat{v}_A(x)v_A(0,z)$ where $v_A(0,z)$ is given by Equation \eqref{eq:alfven_speed_along_z}. This means that resonant absorption can now occur. At the end of the previous section we suggested that the fast wave component of $u_x$ and $u_\perp$ should be negligible near the singularities in resonant absorption experiments and our goal now is to check this claim is valid.

We also change the boundary conditions in $z$ here. We impose periodic boundary conditions in $z$ with period $2L_z$, where the region $0 < z < L_z$ corresponds to the corona and the region $-L_z < z < 0$ corresponds to the chromospehre, with jumps in the background Alfv\'en speed at $z=-L_z$, $z=0$, and $z=L_z$.

\subsection{Normal mode equations}

We seek normal mode solutions and assume the variables are of the form
\begin{equation}
    \label{eq:resonant_abosprtion_general_soln}
    f(\vec{x},t) = f'(x,z)\exp\{i[k_\perp(\cos\alpha\,y - \sin\alpha\,z)+\omega t]\},
\end{equation}
where
\[\omega = \omega_r + i \omega_i.\]
Note that
\[\nabla_\perp f = e^{i[k_\perp(\cos\alpha\,y - \sin\alpha\,z)+\omega t]}\qty(ik_\perp - \sin\alpha\pdv{}{z})f',\]
\[\nabla_{||} f = e^{i[k_\perp(\cos\alpha\,y - \sin\alpha\,z)+\omega t]}\cos\alpha\pdv{f'}{z} .\]
Therefore, Equations \eqref{eq:ux1}-\eqref{eq:b_par1} can be simplified to
\begin{eqnarray}
    i\omega u_x' & = &v_A^2(x,z)\qty[\nabla_{||}'\hat{b}_x' - \pdv{\hat{b}_{||}'}{x}], \\
    i\omega u_\perp' & = & v_A^2(x,z)\qty[\nabla_{||}'\hat{b}_\perp'-\nabla_\perp' \hat{b}_{||}'], \\
    i\omega \hat{b}_x'& = & \nabla_{||}'u_x', \\
    i\omega \hat{b}_\perp' & = & \nabla_{||}'u_{\perp}', \\
    i\omega \hat{b}_{||}' & = & -\qty[\pdv{u_x'}{x}+\nabla_\perp' u_{\perp}'],
\end{eqnarray}
where
\begin{eqnarray}
    \nabla_\perp' & = & \qty(ik_\perp - \sin\alpha\pdv{}{z}), \\
    \nabla_{||}' & = & \cos\alpha\pdv{}{z}.
\end{eqnarray}
Eliminating $\hat{b}_x'$ and $\hat{b}_\perp'$ gives
\begin{eqnarray}
    \label{eq:ux_DAE}
    \pdv{u_x'}{x} & = & -\qty[i\omega \hat{b}_{||'} + \nabla_\perp' u_\perp'], \\
    \label{eq:b_par_DAE}
    \pdv{\hat{b}_{||}'}{x} & = & -\frac{i}{\omega}\mathcal{L}'u_x', \\
    \label{eq:u_perp_DAE}
    \mathcal{L}'u_\perp' & = & i\omega \nabla_\perp' \hat{b}_{||}',
\end{eqnarray}
where
\begin{equation}
    \mathcal{L}'(x,z) = \nabla_{||}'^2+\frac{\omega^2}{v_A^2(x,z)}.
\end{equation}

\subsection{Eigenfunctions and eigenfrequencies}
\label{sec:eigenfunctions_and_eigenfrequencies}

The $z$-domain is given by $-L_z\le z\le L_z$. For mathematical convenience and to help ensure the solutions to Equations \eqref{eq:ux_DAE}-\eqref{eq:u_perp_DAE} are unique we impose periodic boundary conditions in $z$, i.e.
\begin{equation}
    f'(x,z) = f'(x,z+L_z).
\end{equation}
These boundary conditions model a loop going from the chromosphere to the corona then back into the chromosphere and so on. We have a jump in density at $z=-L_z$, $z=0$ and $z=L_z$. The region $0<z<L_z$ corresponds to the corona and the region $-L_z<z<0$ corresponds to the chromosphere. To help solve Equations \eqref{eq:ux_DAE}-\eqref{eq:u_perp_DAE}, we assume the solutions are of the form
\begin{equation}
    \label{eq:eigenfunction_expansion}
    f'(x,z) = \sum_{n=0}^\infty f_n^{(1)}(x)\phi_n(z) + f_n^{(2)}(x)\varphi_n(z),
\end{equation}
where $\phi_n$, $\varphi_n$ are eigenfunctions, with eigenfrequencies, $\omega_n$, $\varpi_n$, respectively. The eigenfunctions satisfy the following Sturm-Liouville equations,
\begin{eqnarray}
    \qty[\nabla_{||}'^2+\frac{\omega_n^2}{v_A^2(0,z)}]\phi_n(z) & = & 0, \\
    \qty[\nabla_{||}'^2+\frac{{\varpi}_n^2}{v_A^2(0,z)}]\varphi_n(z)& = & 0,
\end{eqnarray}
as well as the periodic boundary conditions above. We normalise the eigenfunctions such that
\begin{eqnarray}
    \big\langle \phi_n, \phi_m \big\rangle & = & \frac{v_{A+}^2}{L_z}\int_{-L_z}^{L_z}  \frac{\phi_n \phi_m}{v_A^2(0,z)} dz = \delta_{nm}, \\
    \big\langle \varphi_n, \varphi_m \big\rangle & = & \frac{v_{A+}^2}{L_z}\int_{-L_z}^{L_z}  \frac{\varphi_n \varphi_m}{v_A^2(0,z)} dz = \delta_{nm}, \\
    \big\langle \phi_n, \varphi_m \big\rangle & = & \frac{v_{A+}^2}{L_z}\int_{-L_z}^{L_z}  \frac{\phi_n \varphi_m}{v_A^2(0,z)} dz = 0,
\end{eqnarray}
where $\delta_{nm}$ is the Kronecker delta function. By Sturm-Liouville theory, the eigenfunctions form a basis for the infinite dimensional vector space of twice-differentiable functions satisfying the boundary conditions, this ensures that an expansion of the form \eqref{eq:eigenfunction_expansion} always exists. In Appendix \ref{adx:eigenfunctions_and_eigenfrequencies} we calculate expressions for $\phi_n$, $\varphi_n$, $\omega_n$ and $\varpi_n$.

\subsection{Analytic solution}
\label{sec:analytic_soln}

In this section we aim to calculate the leading order singular/resonant solution to Equations \eqref{eq:ux_DAE}-\eqref{eq:u_perp_DAE} analytically which we will then verify numerically via a graphic approach in Section \ref{sec:numerical_solution}. We calculate the solution by using the method of Frobenius with a similar approach to that used in \citet{Thompson1993,Wright1994,Soler2013}. Note that we only calculate the leading order singular solution here. 

In Appendix \ref{adx:deriving_pde_for_u_perp} we show how $u_x'$ and $\hat{b}_{||}'$ can be eliminated from Equations \eqref{eq:ux_DAE}-\eqref{eq:u_perp_DAE} to give
\begin{equation}
    \tag{\ref{eq:u_perp_2d_order}}
    \mathcal{L}'^2\pdv[2]{u_\perp'}{x}+'\mathcal{L}_x'\mathcal{L}'\pdv{u_\perp'}{x}+\mathcal{M}'u_\perp'=0,
\end{equation}
where
\begin{equation}
    \mathcal{M}' = \mathcal{L}'^3+\mathcal{L}'^2\nabla_\perp'^2+\mathcal{L}_{xx}'\mathcal{L}'-\mathcal{L}_x'^2.
\end{equation}
for $z\ne-L_z$, 0, $L_z$, where $\mathcal{L}'_x(x,z)$ and $a(x)$ are given by Equations \eqref{eq:Lx} and \eqref{eq:a_norm}. 
From the literature (e.g. \citealt{Thompson1993,Wright1996}), we know that resonant absorption can generate singularities in $x$. Equation \eqref{eq:u_perp_2d_order} shows that the coefficient of the leading order $x$-derivative goes to zero if $\mathcal{L}'(x,z)=0$. Let
\[x_{res,m} = x_{r,m} + i x_{i,m},\]
be defined as a location which satisfies
\begin{equation}
    \label{eq:x_res_defn}
    \mathcal{L}'(x_{res,m},z)\phi_m(z)=0.
\end{equation}
We postulate that singularities can form at locations, $x_{res,m}$, which are locations where a resonance can occur.

\subsubsection{Derive equation for \texorpdfstring{$u_{\perp m}^{(1)}(x)$}{uperpm(1)(x)}}

Our goal now is to convert this PDE into an ODE for $u_{\perp m}^{(1)}(x)$ using the expansion given by Equation \eqref{eq:eigenfunction_expansion}. Note that
\begin{eqnarray}
    \label{eq:L_eqn_phi}
    \mathcal{L}'(x,z)\phi_n(z) & = & [\omega^2/\hat{v}_A^2(x) - \omega_n^2]\frac{\phi_n(z)}{v_A^2(0,z)}, \\
    \label{eq:L_eqn_varphi}
    \mathcal{L}'(x,z)\varphi_n(z) & = & [\omega^2/\hat{v}_A^2(x) - \varpi_n^2]\frac{\varphi_n(z)}{v_A^2(0,z)}.
\end{eqnarray}
Multiplying Equation \eqref{eq:u_perp_2d_order} through by $\kappa_m^2(x,z)$ and taking the inner product with $\phi_m(z)$ gives
\begin{equation}
    \label{eq:u_perp_m_1_ode}
    (x-x_{res,m})^2\dv[2]{u_{\perp m}^{(1)}}{x}+(x-x_{res,m})p_m(x)\dv{u_{\perp m}^{(1)}}{x}+q_m(x)u_{\perp m}^{(1)}(x) = f_m(x),
\end{equation}
where
\begin{eqnarray}
    \kappa_m(x,z) & = & (x-x_{res,m})\frac{v_A^2(0,z)}{\omega^2/\hat{v}_A^2(x) - \omega_n^2}, \\
    p_m(x) & = & \mathcal{L}_x'(x,z)\kappa_m(x,z), \\
    q_m(x) & = & \left\langle\mathcal{M}'\phi_m(z), \kappa_m^2(x,z)\phi_m(z)\right\rangle, \\
    f_m(x) & = & -\left\langle\mathcal{M}'u_{\perp m}^{(2)}\varphi_m, \kappa_m^2\phi_m \right\rangle - \sum_{\substack{n=0 \\ n\ne m}}^\infty \left\langle\mathcal{M}'\qty[u_{\perp n}^{(1)}\phi_n + u_{\perp n}^{(2)}\varphi_n],\kappa_m^2\phi_m\right\rangle.
\end{eqnarray}

\subsubsection{Complementary function and particular integral}

To solve Equation \eqref{eq:u_perp_m_1_ode} we will calculate the complementary function first and then the particular integral later. To calculate the complementary function we solve the homogeneous version of Equation \eqref{eq:u_perp_m_1_ode} where $f_m(x)=0$. This can be solved using the method of Frobenius via an approach very similar to that used by \citet{Soler2013} to solve Equation (15) in their paper. By assuming the solutions are of the form
\[u_{\perp m}^{(1)}(x) = (x-x_{res,m})^\sigma\sum_{n=0}^\infty a_n(x-x_{res,m})^n,\]
we can convert the homogeneous version of Equation \eqref{eq:u_perp_m_1_ode} into
\[\sum_{n=0}^\infty a_n[\sigma(\sigma-1)+\sigma p_m(x) + q_m(x)](x-x_{res})^{n+\sigma}.\]
By considering the coefficient of the leading order term, we calculate the indicial equation to be given by
\[I(\sigma)=\sigma(\sigma-1)+p_m(x_{res,m})\sigma + q_m(x_{res,m}).\]
For a non-trivial complementary function we require $I(\sigma)=0$. Since $p_m(x_{res,m})=-q_m(x_{res,m})=1$ we know that $\sigma =\pm1$. Note that the allowed values of $\sigma$ differ by an integer. Therefore, the general complementary solution for $u_{\perp m}^{(1)}(x)$ is given by a linear combination of
\begin{equation}
    u_{\perp m 1}^{(1)}(x) = \sum_{n=1}^\infty a_n (x-x_{res,m})^n,
\end{equation}
and
\begin{equation}
    u_{\perp m 2}^{(1)}(x) = Cu_{\perp m 1}^{(1)}(x) \ln(x-x_{res,m}) + \sum_{n=-1}^\infty b_n(x-x_{res,m})^n,
\end{equation}
where $a_n$, $b_n$, $C$ are constants.

If we know the power series in $x$ for $f_m(x)$ then we can calculate the particular integral to Equation \eqref{eq:u_perp_m_1_ode} by using the method of undetermined coefficients. Since $f_m(x)$ is proportional to $u_{\perp n}^{(1)}(x)$, $u_{\perp n}^{(2)}(x)$, we know that the leading order term can be at most order $(x-x_{res,m})^{-1}$.

\subsubsection{Singular solution}

Our goal  now is to calculate a leading order singular solution for $u_x'$, $u_\perp'$ and $\hat{b}_{||}'$ which we can verify numerically.
Assume that $x_{i,m}\ll R$ where $R$ is the radius of convergence for the power series solution of $u_\perp'(x,z)$ in $x$. We choose the boundary conditions to ensure that to leading order $u_{\perp n}^{(1)}=u_{\perp n}^{(2)}=u_{\perp m}^{(2)}=0$ for all $n\ne m$. This ensures that the leading order solution for $u_{\perp m}(x)$ is given by the leading term of the complementary solution. Let $u_\perp'$ be given by
\begin{equation}
    \label{eq:u_perp_singular_soln}
    u_\perp'(x,z) = u_0\qty[\frac{\phi_m(z)}{X_{m}} + O(X_m\ln X_m,X_m)],
\end{equation}
where we use the boundary conditions to ensure terms of order $X_m^0$ equal zero and $X_m$ is given by
\begin{equation}
    X_m=\frac{x - x_{res,m}}{x_{i,m}}.
\end{equation}

From Equation \eqref{eq:u_perp2} we know that
\begin{equation}
    \label{eq:ux_singular_soln}
    u_x'(x,z) = -u_0\Big[x_{i,m}\ln(X_m)\nabla_\perp'\phi_,(z) + O(X_m,X_m^2\ln X_m)\Big],
\end{equation}
where we impose the boundary condition that terms of order $X_m^0$ equal zero.

From Equation \eqref{eq:u_perp_DAE} we know that to leading order $\hat{b}_{||}'$ satisfies
\begin{equation}
    \label{eq:nabla_perp_b_par_singular_soln}
    \begin{aligned}
   \nabla_\perp' \hat{b}_{||}'(z) &= u_0\frac{\mathcal{L}_x'(x_{r,m},z)x_{i,m}}{i\omega_r}\phi_m(z) + \frac{O(X_m\ln X_m, X_m)}{L_z} \\
   &= 2iu_0\frac{\omega_r x_{i,m}}{a(x_{r,m})v_A^2(x_{r,m},z)}\phi_m(z) + \frac{O(X_m\ln X_m, X_m)}{L_z},
   \end{aligned}
\end{equation}
assuming $\omega_i\ll\omega_r$. We need to solve Equation \eqref{eq:nabla_perp_b_par_singular_soln}. For $\sin\alpha=0$, $\hat{b}_{||}'$ is given by
\begin{equation}
    \hat{b}_{||}'(x,z) = \frac{2u_0\omega_r x_{i,m}}{a(x_{r,m})v_A^2(x_{r,m},z)k_\perp}\phi_m(z) +  O(X_m\ln X_m, X_m).
\end{equation}
For $\sin\alpha\ne0$, we can calculate $b_{||}'(x,z)$ by using an integrating factor. However, if $k_\perp L_z / \sin\alpha = j\pi$ for $j\in\mathds{Z}$ then $\hat{b}_{||}'(x,z)$ is not uniquely defined by the boundary conditions in $z$. To see this, assume $\hat{b}_{||}(x,z)$ is of the form
\[\hat{b}_{||}'(x,z) = \sum_{n=-\infty}^\infty \hat{b}_{||n}(x)\exp(in\frac{\pi}{L_z}z),\]
This ensures that $\hat{b}_{||}(x,z)$ is periodic in $z$ with period $2L_z$. Substituting this into Equation \eqref{eq:nabla_perp_b_par_singular_soln}, multiplying by
$\exp(ij\pi z / L_z) / (2 L_z)$ and integrating from $-L_z$ to $L_z$ gives
\[i\qty(k_\perp - \sin\alpha\, j\frac{\pi}{L_z})\hat{b}_{|| j} = \int_{-L_z}^{L_z}\frac{\hat{b}_{||}'(x,z)}{2L_z}\exp(ij\frac{\pi}{L_z}z)dz.\]
Hence, if $k_\perp L_z / \sin\alpha = j\pi$ for $j\in\mathds{Z}$ then $\hat{b}_{||j}'(x)$ and consequently $\hat{b}_{||}'(x,z)$ is not uniquely defined by the boundary conditions in $z$. 

\subsubsection{Poynting flux}

Finally, we aim to calculate an expression for the Poynting flux because the jump in Poynting flux across the resonant field line describes the rate at which it absorbs energy. The Poynting flux, $\vec{S}$, is given by
\begin{equation}
    \vec{S} = \frac{1}{\mu}\big[(\vec{B}_0\vdot\Re(\vec{b}))\Re(\vec{u}) - (\Re(\vec{u})\vdot\Re(\vec{b}))\vec{B}_0\big],
\end{equation}
where $\Re$ denotes the real part operator.
Hence, the Poynting flux directed in the $x$-direction, $S_x=\vec{S}\vdot\vec{\hat{x}}$ is given by
\begin{equation}
    S_x = \frac{B_0^2}{\mu}\Re\qty(u_x)\Re\qty(\hat{b}_{||}).
\end{equation}
The spatial average in $y$ is given by
\begin{equation}
    \langle S_x \rangle = \frac{B_0^2}{4\mu}\qty(u_x'\hat{b}_{||}'^* + u_x'^*\hat{b}_{||}'),
\end{equation}
where an asterisk, $*$, is used to denote the complex conjugate.
Substituting Equation \eqref{eq:ux_singular_soln} gives
\begin{equation}
    \label{eq:poy_flux_approx}
    \langle S_x \rangle = -\frac{B_0^2}{4\mu}u_0\Big[x_{i,m}\Big(\ln(X_m)\nabla_\perp'\phi_m \hat{b}_{||}'^* + \ln(X_m^*)\nabla_\perp'^*\phi_m\hat{b}_{||}'\Big) + O(X_m, X_m^2\ln X_m)\Big].
\end{equation}
Note that
\[\lim_{x\rightarrow \infty} \ln(X_m) - \lim_{x\rightarrow -\infty} \ln(X_m) = -i\pi\text{sign}(x_{i,m}),\]
\[\lim_{x\rightarrow \infty} \ln(X_m^*) - \lim_{x\rightarrow -\infty} \ln(X_m^*) = i\pi\text{sign}(x_{i,m}),\]
where $\text{sign}(x_{i,m})=x_{i,m}/\abs{x_{i,m}}$.
We define the jump in Poynting flux, $\langle \Delta S_x \rangle$, across $x=x_{r,m}$ as
\begin{equation}
\label{eq:poy_flux_jump}
    \langle \Delta S_x \rangle = i\pi u_0x_{i,m}\frac{B_0^2}{4\mu}\, \text{sign}(x_{i,m})\Big(\nabla_\perp'\phi_n \hat{b}_{||}^* - \nabla_\perp'^*\phi_n \hat{b}_{||}\Big).
\end{equation}

\subsection{Numerical solution}
\label{sec:numerical_solution}

In this section we aim to verify the Equations we derived in Section \ref{sec:analytic_soln} numerically via a graphical approach. To solve the system numerically we use the expansion given by
\eqref{eq:eigenfunction_expansion} to convert Equations \eqref{eq:ux_DAE}-\eqref{eq:u_perp_DAE} from a set of PDEs to a set of ODEs. We then solve the ODEs as an initial value problem in $x$ using \texttt{solve\_ivp} in \citet{Scipy2020}.

For $z\ne0$, $-L_z$, $L_z$, taking $\nabla_\perp$ of Equation \eqref{eq:u_perp_DAE} gives
\begin{equation}
    \label{eq:nabla_perp_u_perp_DAE}
    \mathcal{L}'\nabla_\perp' u_\perp' = i\omega \nabla_\perp'^2 \hat{b}_{||}'.
\end{equation}
Let 
\begin{equation}
    \label{eq:nabla_perp_u_perp_eigenfunction_expansion}
    \nabla_\perp'u_\perp' = \Delta_{\perp0}^{(1)}(x)\phi_0(z) + \sum_{n=1}^\infty \Delta_{\perp n}^{(1)}(x)\phi_n(z) + \Delta_{\perp n}^{(2)}(x)\varphi_n(z).
\end{equation}
Using Equations \eqref{eq:eigenfunction_expansion}, \eqref{eq:nabla_perp_u_perp_eigenfunction_expansion}, \eqref{eq:L_eqn_phi} and \eqref{eq:L_eqn_varphi}, Equations \eqref{eq:ux_DAE}, \eqref{eq:b_par_DAE} and \eqref{eq:nabla_perp_u_perp_DAE} can be written as
\begin{eqnarray}
    \pdv{u_x'}{x} & = & -\sum_{n=0}^\infty i\omega[b_{||n}^{(1)}\phi_n(z)+b_{||n}^{(2)}\varphi_n] + \Delta_{\perp n}^{(1)}\phi_n + \Delta_{\perp n}^{(2)}\varphi_n, \\
    \pdv{b_{||}'}{x} & = & -\frac{i}{\omega}\sum_{n=0}^\infty u_{xn}^{(1)}[\omega^2/\hat{v}_A^2(x) - \omega_n^2]\frac{\phi_n(z)}{v_A^2(0,z)} + u_{xn}^{(2)}[\omega^2/\hat{v}_A^2(x) - \varpi_n^2]\frac{\varphi_n(z)}{v_A^2(0,z)}, \\
    i\omega\nabla_\perp'^2b_{||}' & = & \sum_{n=0}^\infty \Delta_{\perp n}^{(1)}[\omega^2/\hat{v}_A^2(x) - \omega_n^2]\frac{\phi_n(z)}{v_A^2(0,z)}+\Delta_{\perp n}^{(2)}[\omega^2/\hat{v}_A^2(x) - \varpi_n^2]\frac{\varphi_n(z)}{v_A^2(0,z)}.
\end{eqnarray}
Taking the inner product of the above equations with $\phi_m$ and $\varphi_m$ gives
\begin{eqnarray}
    \dv{u_{xm}^{(1)}}{x} & = & -[i\omega b_{||m}^{(1)} + \Delta_{\perp m}^{(1)}], \\
    \dv{u_{xm}^{(2)}}{x} & = & -[i\omega b_{||m}^{(2)} + \Delta_{\perp m}^{(2)}], \\
    \dv{b_{||m}^{(1)}}{x} & = & -\frac{i}{\omega}\sum_{n=0}^\infty u_{xn}^{(1)}[\omega^2/\hat{v}_A^2(x)-\omega_n^2]I_1, \\
    \dv{b_{||m}^{(2)}}{x} & = & -\frac{i}{\omega}\sum_{n=0}^\infty u_{xn}^{(2)}[\omega^2/\hat{v}_A^2(x)-\varpi_n^2]I_2, \\
    \Delta_{\perp m}^{(1)} & = & \frac{i\omega}{\omega^2/\hat{v}_A^2-\omega_m^2}\sum_{n=0}^\infty b_{||n}^{(1)}I_3 + b_{||n}^{(2)}I_4, \\
    \Delta_{\perp m}^{(2)} & = & \frac{i\omega}{\omega^2/\hat{v}_A^2-\varpi_m^2}\sum_{n=0}^\infty  b_{||n}^{(1)}I_5+b_{||n}^{(2)}I_6,
\end{eqnarray}
where
\begin{eqnarray}
    I_1 & = & \left\langle\frac{\phi_n(z)}{v_A^2(0,z)},\phi_m(z)\right\rangle, \\
    I_2 & = & \left\langle\frac{\varphi_n(z)}{v_A^2(0,z)},\varphi_m(z)\right\rangle, \\
    I_3 & = & \big\langle v_A^2(0,z) \nabla_\perp'^2\phi_n(z),\phi_m(z)\big\rangle, \\
    I_4 & = & \big\langle v_A^2(0,z) \nabla_\perp'^2\varphi_n(z),\phi_m(z)\big\rangle, \\
    I_5 & = & \big\langle v_A^2(0,z) \nabla_\perp'^2\phi_n(z),\varphi_m(z)\big\rangle, \\
    I_6 & = & \big\langle v_A^2(0,z) \nabla_\perp'^2\varphi_n(z),\varphi_m(z)\big\rangle.
\end{eqnarray}
This is a set of ODEs which we can solve as an initial value problem in $x$ using \texttt{solve\_ivp} in \citet{Scipy2020}. However, for numerical reasons, only a finite number of terms can be calculated. Therefore, we truncate the summations after a finite number of terms/harmonics, $N_h$.

The Alfv\'en speed $x$-dependence is given by
\begin{equation}
    \hat{v}_A(x)=1+\frac{x}{a_0},
\end{equation}
where $a_0=a(0)$ is a constant which controls the Alfv\'en speed length scale in $x$. From Equation \eqref{eq:x_res_defn}, we can calculate $x_{res,m}$ locations using
\begin{equation}
    \hat{v}_A(x_{res,m}) = \frac{\omega}{\omega_m}.
\end{equation}
This implies that
\begin{eqnarray}
    x_{r,m} & = &  a_0\qty(\frac{\omega_r}{\omega_m}-1), \\
    x_{i,m} & = & a_0\frac{\omega_i}{\omega_m}.
\end{eqnarray}
We choose $v_{A-} = v_{A+} / 21$ somewhat arbitrarily. A larger factor, such as $10^3$, instead of 21 would be more typical in the solar atmosphere. However, a factor of 21 ensures the wavelength in the $z<0$ region is long enough for the shape of the solution to be clearly visible. 21 is also an odd integer which is useful because we have an explicit analytic formula (instead of the implicit formulas given by Equations \ref{eq:omega_n_eqn} and \ref{eq:varpi_n_eqn}) for some of the eigenfrequencies including $\omega_{11}$. We set
\begin{equation}
    \omega_r = \omega_{11}.
\end{equation}
Hence, $x_{r,11}=0$ and $x_{i,11} = a_0\omega_i/\omega_{11}$.

The boundary conditions in $x$ are
\begin{eqnarray}
    \label{eq:ux_x_min_bc}
    u_x'(x_{min},z) & = & -u_0x_{i,11}\ln(X_{11})\nabla_\perp'\phi_{11}(z), \\
    \label{eq:b_par_x_min_bc}
    \nabla_\perp' \hat{b}_{||}'(x_{min},z) & = &2iu_0\frac{\omega_r x_{i,11}}{a_0v_A^2(0,z)}\phi_{11}(z),
\end{eqnarray}
where $k_\perp L_z / \sin\alpha \ne n\pi$ for $n\in\mathds{Z}$ to ensure $\hat{b}_{||}'(x_{min},z)$ is uniquely determined by Equation \eqref{eq:b_par_x_min_bc}.
To calculate $u_{xn}^{(1)}(x_{min})$, $u_{xn}^{(2)}(x_{min})$, $b_{||n}^{(1)}(x_{min})$, $b_{||n}^{(2)}(x_{min})$, for $n=0$, 1, 2, ..., $N_h$ we take the inner product of $u_x'(x_{min},z)$ and $\hat{b}_{||}'(x_{min},z)$ with $\phi_n$ and $\varphi_n$.

\begin{figure*}
    \centering
    \includegraphics[width=17cm,height=0.91\textheight,keepaspectratio]{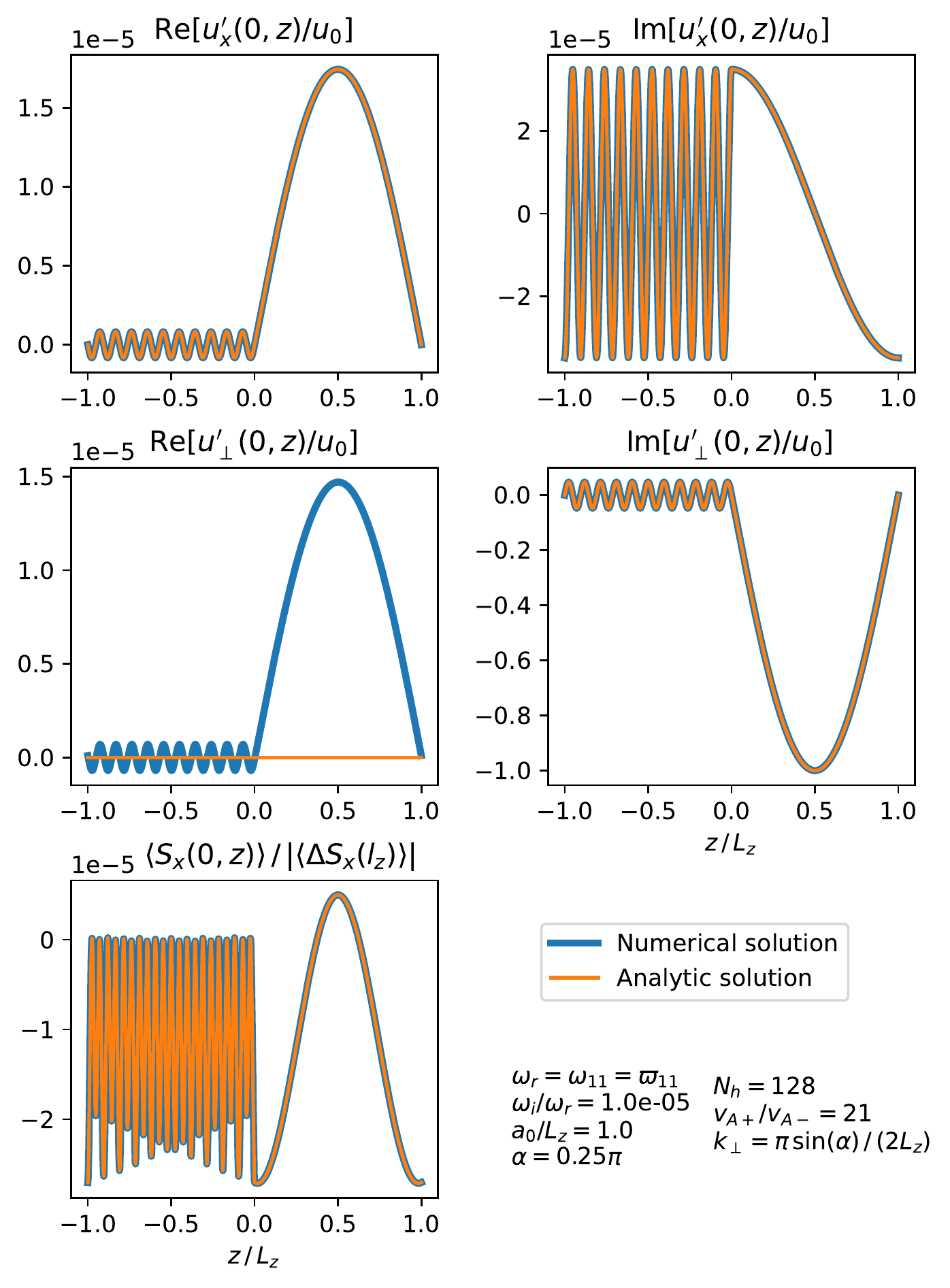}
    \caption{This figure shows plots of the real and imaginary parts of $u_x'$ (top row), $u_\perp'$ (middle row) and $\langle S_x \rangle$ (bottom left) as a function of $z$ at $x=0$. The blue curves were calculated numerically and the orange curves plot the analytic solutions using Equations \eqref{eq:u_perp_singular_soln}, \eqref{eq:ux_singular_soln} and \eqref{eq:poy_flux_approx}. The velocity curves are normalised by, $u_0$, which gives the $u_\perp'$ amplitude at $x=0$. The Poynting flux is normalised by $\abs{\langle \Delta S_x \rangle}$ at $z=l_z$, see Equation \eqref{eq:poy_flux_jump}.}
    \label{fig:along_z}
\end{figure*}

\begin{figure*}
    \centering
    \includegraphics[width=17cm,height=0.91\textheight,keepaspectratio]{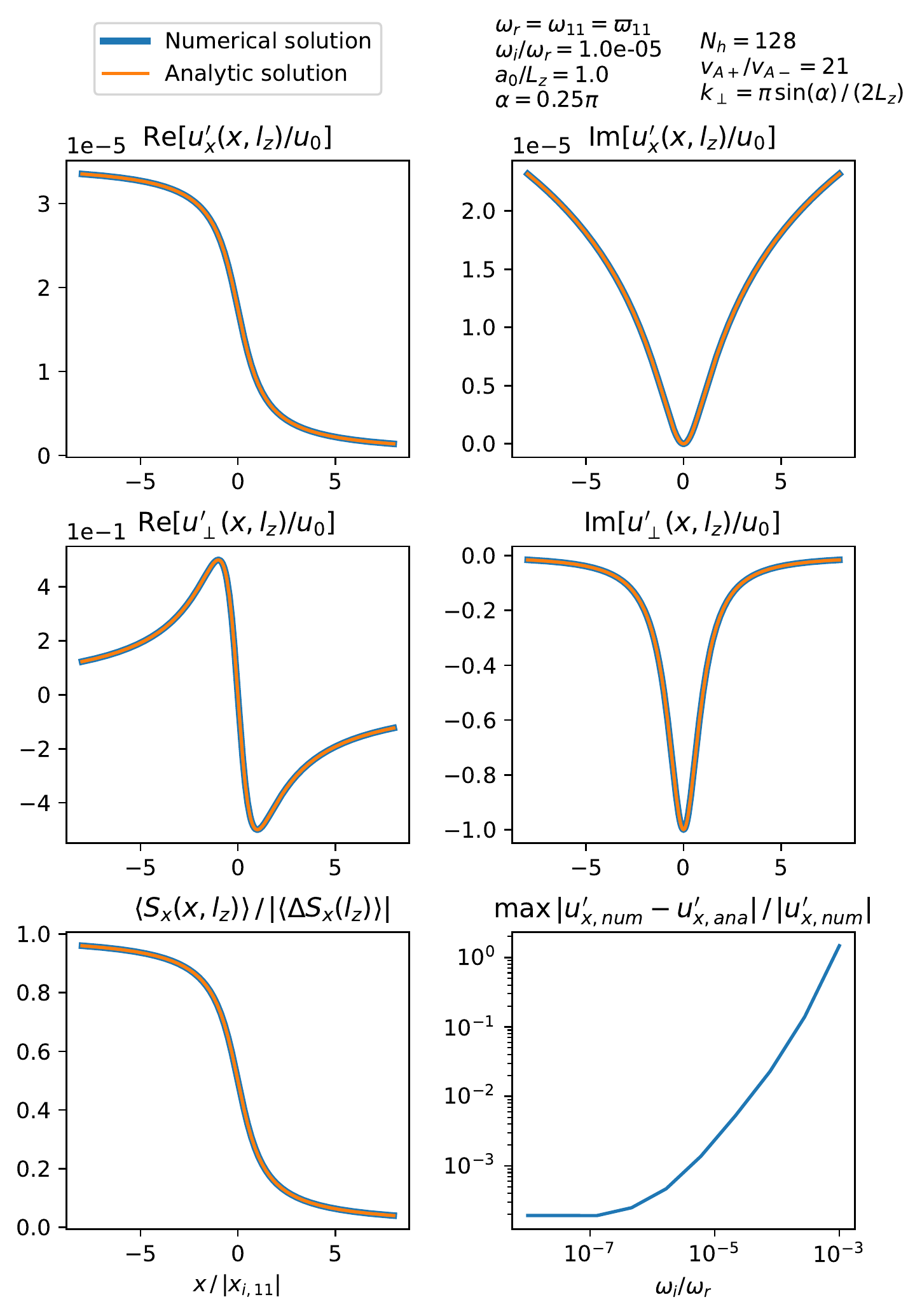}
    \caption{This figure is similar to Figure \ref{fig:along_z} except it plots the variables as function of $x$ at $z=l_z$. The bottom right plot shows the relative error between the numerical and analytic solutions for $u_x$ as a function of $\omega_i / \omega_r$. The solutions are smooth because we use a complex frequency, $\omega$, which moves the singularity/resonant location, $x_{res,11}$, off the real-axis and into the complex plane.}
    \label{fig:along_x}
\end{figure*}

To check that the numerical solutions agree with the analytic solutions given by Equations \eqref{eq:u_perp_singular_soln}, \eqref{eq:ux_singular_soln} and \eqref{eq:poy_flux_approx} and to help visualise the solutions, we plot the solutions in Figure \ref{fig:along_z} and \ref{fig:along_x}. Since $v_{A-} = v_{A+}/21$, the wavelength of the waves in the chromosphere ($z<0$) are shorter by a factor 21 compared with the waves in the corona ($z>0$). They show that the analytic and numerical solutions agree provided $\omega_i/\omega_r$ is small enough. The bottom right of the Figure \ref{fig:along_x} shows the maximum relative error between the numerical and analytic solutions for $u_x$, which is given by
\[
\max_{\substack{-8|x_{i,11}|\le x\le 8|x_{i,11}| \\ -L_z\le z \le L_z}}\abs{\frac{u_{x,num}'-u_{x,ana}'}{u_{x,num}'}},
\]
where $u_{x,num}$ denotes the $u_x$ solution which was calculated numerically and $u_{x,ana}$ was calculated analytically using Equation \eqref{eq:ux_singular_soln}. The bottom right of the Figure \ref{fig:along_x} shows that decreasing $\omega_i/\omega_r$ causes the error to decrease. The plot appears to be converging to a non-zero value, and this is because the numerical solution only uses a finite number of harmonics $N_h$. Increasing $N_h$ acts to reduce the error further. Figure \ref{fig:along_z} shows that although the amplitude of $u_\perp$ may decrease significantly in the chromosphere, this is not necessarily true for $u_x$. This was predicted by Equation \eqref{eq:ux1-_asymptotic_expansion_with_chromosphere}, which showed that even if $v_{A-}\rightarrow0$, $u_x$ in the chromosphere will not go to zero provided the fast waves are evanescent in the chromosphere. This suggests that imposing line-tied ($\vec{u}=0$) boundary conditions at the edge of the corona will lead to the generation of unphysical structures. The usual justification for imposing line-tied boundary conditions is that the chromosphere is significantly denser than the corona and so it acts like a solid wall. Here, $v_{A+}/v_{A-}=21$ and yet $u_x$ is showing no signs of going to zero in the chromosphere. Although $u_x$ does not go to zero, the kinetic energy and magnetic energy is dominated by $u_\perp$ and $b_\perp$. Therefore, the absolute value of the velocity does decrease significantly in the chromosphere.

\newpage

\section{Summary and conclusions}
\label{sec:summary}

The first results presented in this paper showed how Alfv\'en waves propagating from the corona to the chromosphere mode convert at the transition region to form fast waves (see Section \ref{sec:uniform_alfven_speed}). If
\[k_x \ge \Re\qty(\sqrt{k_{||+}^2 - k_y^2}),\]
then the fast waves are evanescent in the corona and will form boundary layers at the transition region (see Figures \ref{fig:piecewise_constant_vs_uniform_real_part} and \ref{fig:piecewise_constant_vs_uniform_imag_part}). This phenomenon has been shown in for example, \citet{Halberstadt1993,Halberstadt1995,Arregui2003}. 
To simulate the conditions near a singularity in a resonant absorption experiment, we calculated asymptotic expansions for $k_x / k_{||+} \rightarrow \infty$ (see Equations \ref{eq:ux1_uniform}-\ref{eq:b_par3_uniform}). They show that the boundary layers do not form if the magnetic field is normal to the transition region, i.e. if $\alpha=0$.

After that, we tested the validity of line-tied boundary conditions by checking if these boundary layers still form if a piecewise constant background Alfv\'en speed with continuity of $\vec{u}$ and $\dv*{\vec{u}}{z}$ is imposed instead (see Section \ref{sec:piecewise_constant_alfven_speed}). We believe we are the first authors to test line-tied boundary conditions in this way. The results (see Figures \ref{fig:piecewise_constant_vs_uniform_real_part}-\ref{fig:fast_wave_error_vs_kx}) suggest that line-tied boundary conditions provide a good approximation provided the fast waves can propagate in the chromosphere, i.e.
\[k_x \le \Re\qty(\sqrt{k_{||-}^2 - k_y^2}).\]
However, if the fast waves are evanescent in the chromosphere, i.e. 
\[k_x \ge \Re\qty(\sqrt{k_{||-}^2 - k_y^2}).\]
then line-tied boundary conditions will overestimate the size of the boundary layers by a factor of the order $k_x / k_{||-}$. If $k_x\gg k_{||-}$ then Equations \eqref{eq:ux1-_asymptotic_expansion_with_chromosphere}-\eqref{eq:b_par3+_asymptotic_expansion_with_chromosphere} approximate the solution, however, if $k_{||+} \ll k_x\ll k_{||-}$ then the line-tied solutions (Equations \ref{eq:ux1_uniform}-\ref{eq:b_par3_uniform}) provide a better approximation. We estimated that if the Alfv\'en waves are able to phase-mix to the shortest length scales allowed by the resistivity and viscosity in the corona then the $k_x\gg k_{||-}$ is usually most appropriate limit. However, during the early stages, before the waves have phase-mixed, $k_x\ll k_{||-}$ is the most valid limit.

The results from Section \ref{sec:piecewise_constant_alfven_speed} suggest that the fast wave component of the velocity should go to zero near singular locations in a resonant absorption experiment since $k_x\rightarrow \infty$ approaching the singularity. In the final section (Section \ref{sec:resonant_absorption_model}), we tested this by calculating the normal mode resonant absorption solution in a domain where the background Alfv\'en speed is a function of $x$ and piecewise constant in $z$. We used periodic boundary conditions in $z$ for convenience and because this simulates a loop which goes from the chromosphere to the corona then back to the chromosphere and so on. In Section \ref{sec:analytic_soln} we calculated the leading order singular solutions near a resonant location analytically. In Section \ref{sec:numerical_solution} we verified the analytic solutions by using eigenfunctions to convert the PDEs into a set of ODEs which were solved numerically. In Figures \ref{fig:along_z} and \ref{fig:along_x} we plot the analytic and numerical solutions together and show that they agree. We verified that to leading order, $u_x$ and $u_\perp$ do not contain any boundary layers near the singularity. This confirms that imposing line-tied boundary conditions can cause the model to significantly overestimate the boundary layers' amplitude. However, in Section \ref{sec:uniform_alfven_speed} we showed as $k_x\rightarrow\infty$ the amplitude of the boundary layers does not increase. This suggests that the boundary layers have a limited effect on the rate at which the resonant absorption occurs. Moreover, since the boundary layers are evanescent, they cannot directly transport energy away.

In conclusion, Alfv\'en waves will (in general) mode convert to form fast waves at the transition region. The fast waves can be evanescent or can propagate depending on the value of $\vec{k}\vdot\vec{k} - \omega^2/v_A^2$. We have demonstrated that line-tied boundary conditions can be more problematic than many authors realise. If $k_x$ grows large enough such that it causes the fast waves in the chromosome to be evanescent, then the line-tied model will fail to reduce the boundary layer's size accurately. We believe that few authors are aware of this because it is quite counter-intuitive that $k_x$ (which gives the length scales in a direction tangential to the chromosphere/corona interface) can be important in determining the validity of line-tied boundary conditions. Increasing $k_x$ can result in the line-tied model greatly overestimating the boundary layer's amplitude by approximately a factor $k_x / k_{||-}$. With that in mind, we believe authors should continue to use line-tied boundary conditions for their simplicity and ability to reduce computation time, provided that their flaws are understood. For example, we have shown that line-tied boundary conditions provide a good approximation (with an error of the order $v_{A-}/v_{A+}$) if the fast waves can propagate in the chromosphere. If the fast waves are evanescent in the chromosphere, then line-tied boundary conditions are less useful. However, sometimes this won't be important. For example, when studying energetics, as the boundary layers are evanescent and cannot transport energy away. Moreover, we showed in Section \ref{sec:uniform_alfven_speed} that their energy does not grow as $k_x\rightarrow \infty$.

\acknowledgments

We thank the referees for their constructive comments. This project has received funding from the Science and Technology Facilities Council (U.K.) through the consolidated grant ST/N000609/1.

\appendix

\section{Asymptotic expansions}
\label{adx:asymptotic_expansions}

\subsection{Uniform background Alfv\'en speed}
\label{adx:uniform_background_alfven_speed}

In this section our goal is to present asymptotic expansions for the coefficients, $u_{xn}$, $u_{\perp n}$, $b_{xn}$, $b_{\perp n}$, $b_{|| n}$, in the model where line-tied boundary conditions are imposed (see Section \ref{sec:uniform_alfven_speed}). To simplify the notation, let 
\begin{equation}
    \hat{k}_y = \frac{k_y}{k_{||+}}.
\end{equation}
To assist with the algebra we use a symbolic computing environment called Maple. In the supplementary material we provide the maple code as well as an exported pdf to enable users without maple installed to view it. By using Equations \eqref{eq:uxn}-\eqref{eq:b_par_u_perp} as well the maple code,
\[\text{maple\_code\_appendix\_A1.mw},\]
which can be accessed in the attached .tar.gz file (with a corresponding pdf file to enable the user read the code without the need for a Maple installation), We calculate that the leading order asymptotic expansions for the $u_x$ coefficients are given by
\begin{eqnarray}
    \label{eq:ux1_uniform}
    \frac{u_{x1}}{u_0} & = & -\frac{\hat{k}_y-\sin\alpha}{\cos\alpha}\frac{k_{||+}}{k_x} + O\qty(\frac{k_{||+}^2}{k_x^2}), \\
    \frac{u_{x2}}{u_0} & = & \frac{\hat{k}_y+\sin\alpha}{\cos\alpha}\frac{k_{||+}}{k_x} + O\qty(\frac{k_{||+}^2}{k_x^2}), \\
    \label{eq:ux3_uniform}
    \frac{u_{x3}}{u_0} & = & -2\tan\alpha\,\frac{k_{||+}}{k_x} + O\qty(\frac{k_{||+}^2}{k_x^2}).
\end{eqnarray}
For large $k_x$, the $u_\perp$ coefficients are given by
\begin{eqnarray}
    \frac{u_{\perp1}}{u_0} & = & 1, \\
    \frac{u_{\perp2}}{u_0} & = & -1 + O\qty(\frac{k_{||+}}{k_x}), \\
    \frac{u_{\perp3}}{u_0} & = & 2i\frac{\sin^2\alpha}{\cos\alpha}\frac{k_{||+}}{k_x} + O\qty(\frac{k_{||+}^2}{k_x^2}).
\end{eqnarray}
To leading order, the $\hat{b}_x$ coefficients are given by
\begin{eqnarray}
    \frac{v_{A+}\hat{b}_{x1}}{u_0} & = & -\frac{\hat{k}_y-\sin\alpha}{\cos\alpha}\frac{k_{||+}}{k_x} + O\qty(\frac{k_{||+}^2}{k_x^2}), \\
    \frac{v_{A+}\hat{b}_{x2}}{u_0} & = & -\frac{\hat{k}_y+\sin\alpha}{\cos\alpha}\frac{k_{||+}}{k_x} + O\qty(\frac{k_{||+}^2}{k_x^2}), \\
    \frac{v_{A+}\hat{b}_{x3}}{u_0} & = & -2i\sin\alpha + O\qty(\frac{k_{||+}}{k_x}).
\end{eqnarray}
Next, the leading order asymptotic expansions for the $\hat{b}_\perp$ coefficients are given by
\begin{eqnarray}
    \frac{v_{A+}\hat{b}_{\perp1}}{u_0} & = & 1, \\
    \frac{v_{A+}\hat{b}_{\perp2}}{u_0} & = & 1 + O\qty(\frac{k_{||+}}{k_x}), \\
    \frac{v_{A+}\hat{b}_{\perp3}}{u_0} & = & -2\sin^2\alpha + O\qty(\frac{k_{||+}}{k_x}).
\end{eqnarray}
Finally, the $\hat{b}_{||}$ coefficients are given by
\begin{eqnarray}
    \frac{v_{A+}\hat{b}_{||1}}{u_0} & = & 0, \\
    \frac{v_{A+}\hat{b}_{||2}}{u_0} & = & 0, \\
    \label{eq:b_par3_uniform}
    \frac{v_{A+}\hat{b}_{||3}}{u_0} & = & \sin(2\alpha) + O\qty(\frac{k_{||+}}{k_x}).
\end{eqnarray}

\subsection{Piecewise constant background Alfv\'en speed}
\label{adx:piecewise_constant_background_alfven_speed}

In this section our goal is to present asymptotic expansions for the coefficients, $u_{xn\pm}$, $u_{\perp n \pm}$, $b_{xn\pm}$, $b_{\perp n\pm}$, $b_{|| n\pm}$, in the model where a pieceiwse constant background Alfv\'en speed is used (see Section \ref{sec:piecewise_constant_alfven_speed}). To simplify the notation, let
\begin{equation}
    r = \frac{k_{||+}}{k_{||-}}.
\end{equation}
Assuming $k_x$ is large, and using Equations \eqref{eq:uxnpm}-\eqref{eq:b_par_pm_u_perp_pm} as well as the maple code, \[\text{maple\_code\_appendix\_A2.mw},\]
which can be be accessed in the attached .tar.gz file, we calculate the leading order asymptotic expansions for the $u_x$ coefficients as
\begin{eqnarray}
    \label{eq:ux1-_asymptotic_expansion_with_chromosphere}
    \frac{u_{x 1-}}{u_0} & = & -2r\frac{r \hat{k}_y-\sin\alpha}{(1+r)\cos\alpha}\frac{k_{||-}}{k_x} + O\qty(\frac{k_{||-}^2}{k_x^2}), \\
    \frac{u_{x 4-}}{u_0} & = & -ir(1-r)\frac{\sin\alpha}{\cos^2\alpha}\frac{k_{||-}^2}{k_x^2} + O\qty(\frac{k_{||-}^3}{k_x^3}), \\
    \frac{u_{x 1+}}{u_0} & = & -r\frac{\hat{k}_y-\sin\alpha}{\cos\alpha}\frac{k_{||-}}{k_x} + O\qty(\frac{k_{||-}^2}{k_x^2}), \\
    \frac{u_{x 2+}}{u_0} & = & r\frac{1-r}{1+r}\frac{\hat{k}_y+\sin\alpha}{\cos\alpha}\frac{k_{||-}}{k_x} + \qty(\frac{k_{||-}^2}{k_x^2}), \\
    \frac{u_{x 3+}}{u_0 } & = & -ir(1-r)\frac{\sin\alpha}{\cos^2\alpha}\frac{k_{||-}^2}{k_x^2} + O\qty(\frac{k_{||-}^3}{k_x^3}).
\end{eqnarray}
The leading order asymptotic expansions for the $u_\perp$ coefficients are given by
\begin{eqnarray}
    \frac{u_{\perp 1-}}{u_0} & = & \frac{2r}{r+1} + O\qty(\frac{k_{||-}^2}{k_x^2}), \\
    \label{eq:u_perp4-}
    \frac{u_{\perp 4-}}{u_0} & = & r(r-1)\tan^2\alpha\frac{k_{||-}^2}{k_x^2} + O\qty(\frac{k_{||-}^3}{k_x^3}), \\
    \frac{u_{\perp 1+}}{u_0} & = & 1, \\
    \frac{u_{\perp 2+}}{u_0} & = & -\frac{1-r}{1+r} + O\qty(\frac{k_{||-}^2}{k_x^2}), \\
    \frac{u_{\perp 3+}}{u_0} & = & -r(1-r)\tan^2\alpha \frac{k_{||-}^2}{k_x^2} + O\qty(\frac{k_{||-}^3}{k_x^3}).
\end{eqnarray}
The leading order asymptotic expansions for the $b_x$ coefficients are given by
\begin{eqnarray}
    \frac{v_{A+}\hat{b}_{x 1-}}{u_0} & = & -2\frac{r\hat{k}_y-\sin\alpha}{(1+r)\cos\alpha}\frac{k_{||-}}{k_x} + O\qty(\frac{k_{||-}^2}{k_x^2}), \\
    \frac{v_{A+}\hat{b}_{x 4-}}{u_0} & = & -(1-r)\tan\alpha\frac{k_{||-}}{k_x} + O\qty(\frac{k_{||-}}{k_x^2}), \\
    \frac{v_{A+}\hat{b}_{x 1+}}{u_0} & = & -r\frac{\hat{k}_y-\sin\alpha}{\cos\alpha}\frac{k_{||-}}{k_x} + O\qty(\frac{k_{||-}^2}{k_x^2}), \\
    \frac{v_{A+}\hat{b}_{x 2+}}{u_0} & = & -r\frac{1-r}{1+r}\frac{\hat{k}_y+\sin\alpha}{\cos\alpha}\frac{k_{||-}}{k_x} + O\qty(\frac{k_{||-}^2}{k_x^2}), \\
    \frac{v_{A+}\hat{b}_{x 3+}}{u_0} & = & (1-r)\frac{k_{||-}}{k_x} + O\qty(\frac{k_{||-}^2}{k_x^2}).
\end{eqnarray}
The leading order asymptotic expansions for the $b_\perp$ coefficients are given by
\begin{eqnarray}
    \frac{v_{A+}\hat{b}_{\perp 1-}}{u_0} & = & \frac{2}{1+r} + O\qty(\frac{k_{||-}^2}{k_x^2}), \\
    \frac{v_{A+}\hat{b}_{\perp 4-}}{u_0} & = & -i(1-r)\frac{\sin^2\alpha}{\cos\alpha}\frac{k_{||-}}{k_x} + O\qty(\frac{k_{||-}^2}{k_x^2}), \\
    \frac{v_{A+}\hat{b}_{\perp 1+}}{u_0} & = & 1, \\
    \frac{v_{A+}\hat{b}_{\perp 2+}}{u_0} & = & \frac{1-r}{1+r} + O\qty(\frac{k_{||-}^2}{k_x^2}), \\
    \frac{v_{A+}\hat{b}_{\perp 3+}}{u_0} & = & -i(1-r)\frac{\sin^2\alpha}{\cos\alpha}\frac{k_{||-}}{k_x} + O\qty(\frac{k_{||-}^2}{k_x^2}).
\end{eqnarray}
Finally, the leading order asymptotic expansions for the $\hat{b}_{||}$ coefficients are given by
\begin{eqnarray}
    \hat{b}_{|| 1-} & = & 0, \\
    \frac{v_{A+}\hat{b}_{|| 4-}}{u_0} & = & ir(1-r)\sin\alpha\frac{k_{||-}}{k_x} + O\qty(\frac{k_{||-}^2}{k_x^2}), \\
    \hat{b}_{|| 1+} & = & 0, \\
    \hat{b}_{|| 2+} & = & 0, \\
    \label{eq:b_par3+_asymptotic_expansion_with_chromosphere}
    \frac{v_{A+}\hat{b}_{|| 3+}}{u_0} & = & ir(1-r)\sin\alpha\frac{k_{||-}}{k_x} + O\qty(\frac{k_{||-}^2}{k_x^2}).
\end{eqnarray}

\section{Check recovery of line-tied equations}
\label{adx:check_recover_of_uniform_equations}

In this section, our goal is to check if we recover the equations we derived in Appendix \ref{adx:uniform_background_alfven_speed} if $k_x \ll k_{||-}$. To simplify the notation, let
\begin{eqnarray}
    \epsilon_- & = & \frac{k_x}{k_{||-}}, \\
    \epsilon_+ & = & \frac{k_{||+}}{k_x}.
\end{eqnarray}
By using Equations \eqref{eq:uxnpm}-\eqref{eq:b_par_pm_u_perp_pm} as well as the maple code, \[\text{maple\_code\_appendix\_B.mw},\]
which can be be accessed in the attached .tar.gz file,
we calculate the leading order multivariate Taylor expansions (assuming $\epsilon_-,\epsilon_+\ll1$). The $u_x$ coefficients are given by
\begin{eqnarray}
    \frac{u_{x1-}}{u_0} & = & -2i\sin\alpha\,\epsilon_-\epsilon_+ + O(\epsilon_-^3,\epsilon_-^2\epsilon_+,\epsilon_-\epsilon_+^2,\epsilon_+^3), \\
    \frac{u_{x4-}}{u_0} & = & -2\frac{\cos\alpha}{\sin\alpha}\epsilon_-^2\epsilon_+ + O(\epsilon_-^4,\epsilon_-^3\epsilon_+,...\,,\epsilon_+^4), \\
    \frac{u_{x1+}}{u_0} & = & -\frac{\hat{k}_y - \sin\alpha}{\cos\alpha}\epsilon_+ + O(\epsilon_-^2,\epsilon_-\epsilon_+,\epsilon_+^2), \\
    \frac{u_{x2+}}{u_0} & = & \frac{\hat{k}_y + \sin\alpha}{\cos\alpha}\epsilon_+ + O(\epsilon_-^2,\epsilon_-\epsilon_+,\epsilon_+^2), \\
    \frac{u_{x3+}}{u_0} & = & -2\tan\alpha\,\epsilon_+ + O(\epsilon_-^2,\epsilon_-\epsilon_+,\epsilon_+^2).
\end{eqnarray}
The $u_\perp$ coefficients are given by
\begin{eqnarray}
    \frac{u_{\perp1-}}{u_0}& = &-2i\cos\alpha\,\epsilon_-^2\epsilon_+ + O(\tiny{\epsilon_-^4,\epsilon_-^3\epsilon_+,...\,,\epsilon_+^4}) \\
    \frac{u_{\perp4-}}{u_0} & = & 2\cos\alpha\,\epsilon_-\epsilon_++O(\epsilon_-^2,\epsilon_-\epsilon_+,\epsilon_+^2), \\
    \frac{u_{\perp 1+}}{u_0} & = & 1, \\
    \frac{u_{\perp 2+}}{u_0} & = & -1 + O(\epsilon_-,\epsilon_+), \\
    \frac{u_{\perp 3+}}{u_0} & = & 2i\frac{\sin^2\alpha}{\cos\alpha}\epsilon_+ + O(\epsilon_-^2,\epsilon_-\epsilon_+,\epsilon_+^2).
\end{eqnarray}
The $b_x$ coefficients are given by
\begin{eqnarray}
    \frac{v_{A+}\hat{b}_{x1-}}{u_0} & = & -2i\sin\alpha + O(\epsilon_-,\epsilon_+), \\
    \frac{v_{A+}\hat{b}_{x4-}}{u_0} & = & -2\frac{\cos^2\alpha}{\sin\alpha}\epsilon_- + O(\epsilon_-^2,\epsilon_-\epsilon_+,\epsilon_+^2), \\
    \frac{v_{A+}\hat{b}_{x1+}}{u_0} & = & -\frac{\hat{k}_y - \sin\alpha}{\cos\alpha}\epsilon_+ + O(\epsilon_-^2,\epsilon_-\epsilon_+,\epsilon_+^2), \\
    \frac{v_{A+}\hat{b}_{x2+}}{u_0} & = & -\frac{\hat{k}_y+\sin\alpha}{\cos\alpha}\epsilon_+ + O(\epsilon_-^2,\epsilon_-\epsilon_+,\epsilon_+^2), \\
    \frac{v_{A+}\hat{b}_{x3+}}{u_0} & = & -2i\sin\alpha+O(\epsilon_-,\epsilon_+).
\end{eqnarray}
The $b_\perp$ coefficients are given by
\begin{eqnarray}
    \frac{v_{A+}b_{\perp1-}}{u_0} & = & -2i\cos\alpha\,\epsilon_- + O(\epsilon_-^2,\epsilon_-\epsilon_+,\epsilon_+^2), \\
    \frac{v_{A+}b_{\perp4-}}{u_0} & = & 2\cos^2\alpha + O(\epsilon_-,\epsilon_+), \\
    \frac{v_{A+}b_{\perp1+}}{u_0} & = & 1, \\
    \frac{v_{A+}b_{\perp2+}}{u_0} & = & 1 + O(\epsilon_-,\epsilon_+), \\
    \frac{v_{A+}b_{\perp3+}}{u_0} & = & -2\sin^2\alpha + O(\epsilon_-, \epsilon_+).
\end{eqnarray}
Finally,  the $\hat{b}_{||}$ coefficients are given by
\begin{eqnarray}
    \frac{v_{A+}b_{||1-}}{u_0} & = & 0, \\
    \frac{v_{A+}b_{||4-}}{u_0} & = & \sin(2\alpha) + O(\epsilon_-,\epsilon_+), \\
    \frac{v_{A+}b_{||1+}}{u_0} & = & 0, \\
    \frac{v_{A+}b_{||2+}}{u_0} & = & 0, \\
    \frac{v_{A+}b_{||3+}}{u_0} & = & \sin(2\alpha) + O(\epsilon_-, \epsilon_+).
\end{eqnarray}
Taking the limit $\epsilon_-\rightarrow0$ we do indeed recover the equations derived in Appendix \ref{adx:uniform_background_alfven_speed} for $z>0$.

\section{Eigenfunctions and eigenfrequencies}
\label{adx:eigenfunctions_and_eigenfrequencies}

In this section our goal is to calculate expressions for the eigenfunctions and corresponding eigenfrequencies, $\phi_n(z)$, $\varphi_n(z)$, $\omega_n$, $\varpi_n$, which were discussed in Section \ref{sec:eigenfunctions_and_eigenfrequencies}. By symmetry, we assume the eigenfunctions are of the form
\begin{eqnarray}
\phi_n(z) & = & A_n\begin{cases}
a_n\cos[k_{zn-} (z+l_z)], & z < 0, \\
c_n\cos[k_{zn+} (z-l_z)], & z \ge 0, \\
\end{cases} \\
\varphi_n(z) & = & B_n\begin{cases}
b_n\sin[\bar{k}_{zn-} (z+l_z)], & z < 0, \\
d_n\sin[\bar{k}_{zn+} (z-l_z)], & z \ge 0, \\
\end{cases}
\end{eqnarray}
where $a_n$, $b_n$, $c_n$, $d_n$ are constants, $L_z = 2l_z$ and
\begin{eqnarray}
    k_{zn\pm} & = & \frac{\omega_n}{v_{A\pm}\cos\alpha}, \\
    \bar{k}_{zn\pm} & = & \frac{\varpi_n}{v_{A\pm}\cos\alpha}.
\end{eqnarray}
We require continuity of $\phi_n$, $\varphi_n$, $\dv*{\phi_n}{z}$, $\dv*{\varphi_n}{z}$ at $z=0$ and $z=\pm L_z$, this gives the following equations
\[\begin{aligned}
\begin{pmatrix}
\cos(k_{zn-}l_z) & -\cos(k_{zn+}l_z) \\
k_{zn-}\sin(k_{zn-}l_z) & k_{zn+}\sin(k_{zn+}l_z)
\end{pmatrix}
\begin{pmatrix}
a_n \\
c_n
\end{pmatrix}
&=
0, \\
\begin{pmatrix}
\sin(\bar{k}_{zn-}l_z) & \sin(\bar{k}_{zn+}l_z) \\
\bar{k}_{zn-}\cos(\bar{k}_{zn-}l_z) & -\bar{k}_{zn+}\cos(\bar{k}_{zn+}l_z)
\end{pmatrix}
\begin{pmatrix}
b_n \\
d_n
\end{pmatrix}
&=
0.
\end{aligned}\]
For non-trivial solutions to exist, $\omega_n$, $\varpi_n$ must satisfy
\begin{eqnarray}
\label{eq:omega_n_eqn}
k_{zn+}\sin(k_{zn+}l_z)\cos(k_{zn-}l_z)+ k_{zn-}\sin(k_{zn-}l_z)\cos(k_{zn+}l_z) &=& 0, \\
\label{eq:varpi_n_eqn}
\bar{k}_{zn+}\sin(\bar{k}_{zn-}l_z)\cos(\bar{k}_{zn+}l_z)+\bar{k}_{zn-}\sin(\bar{k}_{zn+}l_z)\cos(\bar{k}_{zn-}l_z) & = & 0.
\end{eqnarray}
These equations define, $\omega_n$, $\varpi_n$ and we order them such that
\begin{equation}
    \omega_0<\omega_1<\omega_2...,\quad \varpi_0<\varpi_1<\varpi_2...\,,
\end{equation}
where $\omega_0=\varpi_0=0$.

If $\omega_n$, $\varpi_n$ satisfy Equations \eqref{eq:omega_n_eqn} and \eqref{eq:varpi_n_eqn}, then the boundary conditions and continuity conditions are ensured if
\begin{eqnarray}
a_n & = & \begin{cases}
\cos(k_{zn+} l_z), & \cos(k_{zn+} l_z) \ne 0, \\
k_{zn+}\sin(k_{zn+} l_z), & \cos(k_{zn+} l_z) = 0, \\
\end{cases} \\
c_n & = & \begin{cases}
\cos(k_{zn-} l_z), & \cos(k_{zn-} l_z) \ne 0, \\
-k_{zn-}\sin(k_{zn-} l_z), & \cos(k_{zn-} l_z) = 0, \\
\end{cases} \\
b_n & = & \begin{cases}
\sin(\bar{k}_{zn+} l_z), & \sin(\bar{k}_{zn+} l_z) \ne 0, \\
\bar{k}_{zn+}\cos(\bar{k}_{zn+} l_z), & \sin(\bar{k}_{zn+} l_z) = 0, \\
\end{cases} \\
d_n & = & \begin{cases}
-\sin(\bar{k}_{zn-} l_z), & \sin(\bar{k}_{zn-} l_z) \ne 0, \\
\bar{k}_{zn-}\cos(\bar{k}_{zn-} l_z), & \sin(\bar{k}_{zn-} l_z) = 0. \\
\end{cases}
\end{eqnarray}

We normalise such that
\begin{equation}
    \Big\langle \phi_n(z), \phi_m(z) \Big\rangle = \Big\langle \varphi_n(z), \varphi_m(z) \Big\rangle = \delta_{nm},
\end{equation}
where $\delta_{nm}$ is the Kronecker delta.
Note that
\begin{eqnarray}
    \langle \phi_n, \phi_n \rangle & = & \frac{v_{A+}^2A_n^2}{L_z}\qty{\frac{a_n^2}{2v_{A-}^2k_{zn-}}\Big[2k_{zn-}l_z+\sin(2k_{zn-}l_z)\Big] + \frac{c_n^2}{2v_{A+}^2k_{zn+}}\Big[2k_{zn+}l_z+\sin(2k_{zn+}l_z)\Big]}, \\
    \langle \varphi_n, \varphi_n \rangle & = & \frac{v_{A+}^2B_n^2}{L_z} \qty{\frac{b_n^2}{2v_{A-}^2\bar{k}_{zn-}}\Big[2\bar{k}_{zn-}l_z-\sin(2\bar{k}_{zn-}l_z)\Big] + \frac{d_n^2}{2v_{A+}^2\bar{k}_{zn-}}\Big[2\bar{k}_{zn-}l_z-\sin(2\bar{k}_{zn-}l_z)\Big]}.
\end{eqnarray}
Hence,
\begin{eqnarray}
    A_n &=& \qty(\frac{v_{A+}^2}{L_z}\qty{\frac{a_n^2}{2v_{A-}^2k_{zn-}}\Big[2k_{zn-}l_z+\sin(2k_{zn-}l_z)\Big] + \frac{c_n^2}{2v_{A+}^2k_{zn+}}\Big[2k_{zn+}l_z+\sin(2k_{zn+}l_z)\Big]})^{-1/2}, \\
    B_n &=& \qty(\frac{v_{A+}^2}{L_z} \qty{\frac{b_n^2}{2v_{A-}^2\bar{k}_{zn-}}\Big[2\bar{k}_{zn-}l_z-\sin(2\bar{k}_{zn-}l_z)\Big] + \frac{d_n^2}{2v_{A+}^2\bar{k}_{zn-}}\Big[2\bar{k}_{zn-}l_z-\sin(2\bar{k}_{zn-}l_z)\Big]})^{-1/2}.
\end{eqnarray}

\section{Deriving equation for \texorpdfstring{$u_\perp'$}{uperp'}}
\label{adx:deriving_pde_for_u_perp}

Eliminating $\hat{b}_{||}'$ from Equations \eqref{eq:ux_DAE}-\eqref{eq:u_perp_DAE} gives
\begin{eqnarray}
    \label{eq:ux2}
    \qty[\mathcal{L}'+\pdv[2]{}{x}]u_x' & = & -\pdv{}{x}\nabla_\perp' u_\perp', \\
    \label{eq:u_perp2}
    [\mathcal{L}'+\nabla_\perp^2]u_\perp'& = &-\pdv{}{x}\nabla_\perp' u_x' 
\end{eqnarray}
We can eliminate $u_x'$ with the following procedure. 
Take the $x$-derivative of the Equation \eqref{eq:ux2}
\[\mathcal{L}_x'u_x'+\qty[\mathcal{L}'+\pdv[2]{}{x}]\pdv{u_x'}{x}=-\nabla_\perp\pdv[2]{u_\perp'}{x},\]
\[\implies u_x'=-\frac{1}{\mathcal{L}_x'}\qty{\qty[\mathcal{L'}+\pdv[2]{}{x}]\pdv{u_x'}{x}+\nabla_\perp'\pdv[2]{u_\perp'}{x}}.\]
where
\begin{eqnarray}
    \label{eq:Lx}
    \mathcal{L}_x' & = & \pdv{\mathcal{L}'}{x}=\frac{-2\omega^2}{a(x)v_A^2(x,z)}, \\
    \label{eq:a_norm}
    a(x) & = & \frac{\hat{v}_A(x)}{\dv*{\hat{v}_A}{x}}.    
\end{eqnarray}
Substitute this into Equation \eqref{eq:ux1} to give
\[-\frac{\mathcal{L}'}{\mathcal{L}_x'}\qty{\qty[\mathcal{L}'+\pdv[2]{}{x}]\pdv{u_x'}{x}+\nabla_\perp'\pdv[2]{u_\perp'}{x}}+\pdv[2]{u_x'}{x}=-\nabla_\perp'\pdv{u_\perp'}{x},\]
Applying $-\nabla_\perp'$ to both sides, substituting for $u_x'$ and assuming $z\ne0,-L_z,L_z$ gives
\[-\frac{\mathcal{L}'}{\mathcal{L}_x'}\qty{\qty[\mathcal{L}'+\pdv[2]{}{x}][\mathcal{L}'+\nabla_\perp'^2]u_\perp'-\nabla_\perp'^2\pdv[2]{u_\perp'}{x}}+\pdv{}{x}[\mathcal{L}'+\nabla_\perp'^2]u_\perp'=\nabla_\perp'^2\pdv{u_\perp'}{x}.\]
This can be simplified to give
\begin{equation}
    \label{eq:u_perp_2d_order}
    \mathcal{L}'^2\pdv[2]{u_\perp'}{x}+'\mathcal{L}_x'\mathcal{L}'\pdv{u_\perp'}{x}+(\mathcal{L}'^3+\mathcal{L}'^2\nabla_\perp'^2+\mathcal{L}_{xx}'\mathcal{L}'-\mathcal{L}_x'^2)u_\perp'=0.
\end{equation}

\newpage

\bibliography{bibliography}{}

\begin{thebibliography}{}
\expandafter\ifx\csname natexlab\endcsname\relax\def\natexlab#1{#1}\fi
\providecommand{\url}[1]{\href{#1}{#1}}
\providecommand{\dodoi}[1]{doi:~\href{http://doi.org/#1}{\nolinkurl{#1}}}
\providecommand{\doeprint}[1]{\href{http://ascl.net/#1}{\nolinkurl{http://ascl.net/#1}}}
\providecommand{\doarXiv}[1]{\href{https://arxiv.org/abs/#1}{\nolinkurl{https://arxiv.org/abs/#1}}}

\bibitem[{{Antolin} {et~al.}(2016){Antolin}, {De Moortel}, {Van Doorsselaere},
  \& {Yokoyama}}]{Antolin2016}
{Antolin}, P., {De Moortel}, I., {Van Doorsselaere}, T., \& {Yokoyama}, T.
  2016, \apjl, 830, L22, \dodoi{10.3847/2041-8205/830/2/L22}

\bibitem[{{Arregui}(2015)}]{Arregui2015}
{Arregui}, I. 2015, Philosophical Transactions of the Royal Society of London
  Series A, 373, 20140261, \dodoi{10.1098/rsta.2014.0261}

\bibitem[{{Arregui} {et~al.}(2003){Arregui}, {Oliver}, \&
  {Ballester}}]{Arregui2003}
{Arregui}, I., {Oliver}, R., \& {Ballester}, J.~L. 2003, \aap, 402, 1129,
  \dodoi{10.1051/0004-6361:20030312}

\bibitem[{{Cally} \& {Hansen}(2011)}]{Cally2011}
{Cally}, P.~S., \& {Hansen}, S.~C. 2011, \apj, 738, 119,
  \dodoi{10.1088/0004-637X/738/2/119}

\bibitem[{{Cargill} {et~al.}(2016){Cargill}, {De Moortel}, \&
  {Kiddie}}]{Cargill2016}
{Cargill}, P.~J., {De Moortel}, I., \& {Kiddie}, G. 2016, \apj, 823, 31,
  \dodoi{10.3847/0004-637X/823/1/31}

\bibitem[{{Cranmer}(2018)}]{Cranmer2018}
{Cranmer}, S.~R. 2018, \apj, 862, 6, \dodoi{10.3847/1538-4357/aac953}

\bibitem[{{Cranmer} \& {van Ballegooijen}(2005)}]{Cranmer2005}
{Cranmer}, S.~R., \& {van Ballegooijen}, A.~A. 2005, \apjs, 156, 265,
  \dodoi{10.1086/426507}

\bibitem[{{De Moortel} \& {Nakariakov}(2012)}]{DeMoortel2012}
{De Moortel}, I., \& {Nakariakov}, V.~M. 2012, Philosophical Transactions of
  the Royal Society of London Series A, 370, 3193,
  \dodoi{10.1098/rsta.2011.0640}

\bibitem[{{Goedbloed} \& {Halberstadt}(1994)}]{Goedbloed1994}
{Goedbloed}, J.~P., \& {Halberstadt}, G. 1994, \aap, 286, 275

\bibitem[{{Goossens} {et~al.}(2011){Goossens}, {Erd{\'e}lyi}, \&
  {Ruderman}}]{Goossens2011}
{Goossens}, M., {Erd{\'e}lyi}, R., \& {Ruderman}, M.~S. 2011, \ssr, 158, 289,
  \dodoi{10.1007/s11214-010-9702-7}

\bibitem[{{Halberstadt} \& {Goedbloed}(1993)}]{Halberstadt1993}
{Halberstadt}, G., \& {Goedbloed}, J.~P. 1993, \aap, 280, 647

\bibitem[{{Halberstadt} \& {Goedbloed}(1995)}]{Halberstadt1995}
---. 1995, \aap, 301, 559

\bibitem[{{Hansen} \& {Cally}(2012)}]{Hansen2012}
{Hansen}, S.~C., \& {Cally}, P.~S. 2012, \apj, 751, 31

\bibitem[{{Heyvaerts} \& {Priest}(1983)}]{Heyvaerts1983}
{Heyvaerts}, J., \& {Priest}, E.~R. 1983, \aap, 117, 220

\bibitem[{{Hollweg}(1984)}]{Hollweg1984}
{Hollweg}, J.~V. 1984, \apj, 277, 392, \dodoi{10.1086/161706}

\bibitem[{{Howson} {et~al.}(2017){Howson}, {De Moortel}, \&
  {Antolin}}]{Howson2017}
{Howson}, T.~A., {De Moortel}, I., \& {Antolin}, P. 2017, \aap, 607, A77,
  \dodoi{10.1051/0004-6361/201731178}

\bibitem[{{Ionson}(1982)}]{Ionson1982}
{Ionson}, J.~A. 1982, \apj, 254, 318

\bibitem[{{Klimchuk}(2015)}]{Klimchuk2015}
{Klimchuk}, J.~A. 2015, Philosophical Transactions of the Royal Society of
  London Series A, 373, 20140256, \dodoi{10.1098/rsta.2014.0256}

\bibitem[{{McIntosh} \& {De Pontieu}(2012)}]{McIntosh2012}
{McIntosh}, S.~W., \& {De Pontieu}, B. 2012, \apj, 761, 138,
  \dodoi{10.1088/0004-637X/761/2/138}

\bibitem[{{McIntosh} {et~al.}(2011){McIntosh}, {de Pontieu}, {Carlsson},
  {Hansteen}, {Boerner}, \& {Goossens}}]{McIntosh2011}
{McIntosh}, S.~W., {de Pontieu}, B., {Carlsson}, M., {et~al.} 2011, \nat, 475,
  477, \dodoi{10.1038/nature10235}

\bibitem[{{Nakariakov} {et~al.}(1999){Nakariakov}, {Ofman}, {Deluca},
  {Roberts}, \& {Davila}}]{Nakariakov1999}
{Nakariakov}, V.~M., {Ofman}, L., {Deluca}, E.~E., {Roberts}, B., \& {Davila},
  J.~M. 1999, Science, 285, 862, \dodoi{10.1126/science.285.5429.862}

\bibitem[{{Priest}(2014)}]{Priest2014}
{Priest}, E. 2014, {Magnetohydrodynamics of the Sun} (Cambridge University
  Press)

\bibitem[{{Prokopyszyn} \& {Hood}(2019)}]{Prokopyszyn2019}
{Prokopyszyn}, A.~P.~K., \& {Hood}, A.~W. 2019, \aap, 632, A93,
  \dodoi{10.1051/0004-6361/201936658}

\bibitem[{{Ruderman} \& {Roberts}(2002)}]{Ruderman2002}
{Ruderman}, M.~S., \& {Roberts}, B. 2002, \apj, 577, 475,
  \dodoi{10.1086/342130}

\bibitem[{{Soler} {et~al.}(2013){Soler}, {Goossens}, {Terradas}, \&
  {Oliver}}]{Soler2013}
{Soler}, R., {Goossens}, M., {Terradas}, J., \& {Oliver}, R. 2013, \apj, 777,
  158, \dodoi{10.1088/0004-637X/777/2/158}

\bibitem[{{Terradas} {et~al.}(2006){Terradas}, {Oliver}, \&
  {Ballester}}]{Terradas2006}
{Terradas}, J., {Oliver}, R., \& {Ballester}, J.~L. 2006, \apj, 642, 533,
  \dodoi{10.1086/500730}

\bibitem[{{Thompson} \& {Wright}(1993)}]{Thompson1993}
{Thompson}, M.~J., \& {Wright}, A.~N. 1993, \jgr, 98, 15541,
  \dodoi{10.1029/93JA00791}

\bibitem[{{Tomczyk} {et~al.}(2007){Tomczyk}, {McIntosh}, {Keil}, {Judge},
  {Schad}, {Seeley}, \& {Edmondson}}]{Tomczyk2007}
{Tomczyk}, S., {McIntosh}, S.~W., {Keil}, S.~L., {et~al.} 2007, Science, 317,
  1192, \dodoi{10.1126/science.1143304}

\bibitem[{Virtanen {et~al.}(2020)Virtanen, Gommers, Oliphant, Haberland, Reddy,
  Cournapeau, Burovski, Peterson, Weckesser, Bright, {van der Walt}, Brett,
  Wilson, Millman, Mayorov, Nelson, Jones, Kern, Larson, Carey, Polat, Feng,
  Moore, {VanderPlas}, Laxalde, Perktold, Cimrman, Henriksen, Quintero, Harris,
  Archibald, Ribeiro, Pedregosa, {van Mulbregt}, \& {SciPy 1.0
  Contributors}}]{Scipy2020}
Virtanen, P., Gommers, R., Oliphant, T.~E., {et~al.} 2020, Nature Methods, 17,
  261, \dodoi{10.1038/s41592-019-0686-2}

\bibitem[{{Wright} \& {Allan}(1996)}]{Wright1996}
{Wright}, A.~N., \& {Allan}, W. 1996, \jgr, 101, 17399,
  \dodoi{10.1029/96JA01141}

\bibitem[{{Wright} \& {Thompson}(1994)}]{Wright1994}
{Wright}, A.~N., \& {Thompson}, M.~J. 1994, Physics of Plasmas, 1, 691,
  \dodoi{10.1063/1.870815}

\end{thebibliography}
\bibliographystyle{aasjournal}

\end{document}